\documentclass[a4paper,12pt]{article}
\usepackage[affil-it]{authblk}
\usepackage{cite}
\usepackage{amsmath}
\usepackage{mathtools}
\usepackage{graphicx}
\usepackage{xcolor}
\usepackage[cp1252]{inputenc}
\usepackage{float}
\usepackage{mathtools}
\usepackage{tabu}
\usepackage{amssymb}
\usepackage{tikz}

\providecommand{\keywords}[1]
{
  \small	
  \textbf{\textit{Keywords:}} #1
}
\usepackage{tikz}
\makeatletter
\DeclareRobustCommand*{\bfseries}{%
   \not@math@alphabet\bfseries\mathbf
   \fontseries\bfdefault\selectfont
   \boldmath
}
\makeatother
\usepackage[square,numbers,sort&compress]{natbib}
\usepackage{tikz}
\usetikzlibrary{arrows}
\usepackage[skip=60pt]{caption}
\usepackage{subcaption}
\usepackage{xfp}
\usepackage{numerica}

\begin{document}
\title{ The Impact  of Thermosolutal Convection on Melting Dynamics of  Nano-enhanced Phase Change Materials (NePCM) }
\author[1,2]{Yousef M. F. El Hasadi \footnote{Email address for correspondence: yme0001@auburn.edu,g.damianidis.al.chasanti@tue.nl}}
\affil[1]{Eindhoven University of Technology, Mechanical Engineering Department, The Netherlands}
\affil[2]{Process and Energy department, Delft University of Technology,Leeghwaterstraat 39,2628 CB Delft, the Netherlands}
\date{}
\maketitle
\begin{abstract}
Nanoparticle-Enhanced Phase Change Materials (NePCM) have been a subject of intensive research owing to their potential for enhanced thermo-physical properties. However, their behavior during phase change processes, such as melting or solidification, remains inadequately understood\@. This investigation focuses on the melting process of NePCM in a square cavity, exploring distinct cases of melting from both the top and bottom sides. The NePCM comprises copper nanoparticles (2 nm in size) suspended in water. Our study involves different combinations of constant temperature boundary conditions and particle volume fractions\@. Utilizing a numerical model based on the one-fluid mixture approach combined with the single-domain enthalpy-porosity model, we account for the phase change process and particles' interaction with the solid-liquid interface. When melting NePCM from the top side, convection effects are suppressed, resulting in a melting process primarily governed by conduction. Both NePCM and pure water melt at the same rate under these conditions\@. However, melting NePCM from the bottom side induces convection-dominated melting. For pure water, thermal convection leads to the formation of convection cells during melting. Contrastingly, melting NePCM triggers thermosolutal convection due to temperature and particle concentration gradients. The flow cells formed from thermosolutal convection in NePCM differ from those in pure water driven by pure thermal convection\@. Our simulations reveal that thermosolutal convection contributes to decelerating the solid-liquid interface, thereby prolonging NePCM melting compared to pure water. Surprisingly, the viscosity increase in NePCM plays a minimal role in the deceleration process, contrary to prior literature attributing slow-downs of the melting process of the NePCM primarily to increased viscosity\@.
\end{abstract}
\\

\section{Introduction}
In light of the substantial harm caused by climate change and the escalation of global temperatures, numerous countries are actively pursuing the decarbonization of their economies. Within modern economies, heat production plays a pivotal role, where heating and cooling activities contribute to 50\% of global energy consumption and 45\% of total worldwide carbon emissions \cite{LDES}. Thermal energy storage (TES) can be essential in decarbonising the heat production process\@. Thermal energy storage can be achieved mainly by two ways the first is sensible heat storage, and the second is latent heat storage \cite{sharma2009review}\@. Latent heat storage is achieved using what we call phase change materials (PCM), which they absorb, and realise a  substantial amount of heat during their phase change process \cite{nazir2019recent}\@. PCMs could be organic, non-organic, or eutectic mixtures \cite{nazir2019recent}\@. Latent heat storage have substantial number of applications such as energy storage in concentrated solar power, thermal management, cold storage, thermal storage in buildings, district heating systems, smart thermal grid, and thermal management \citep{aldoss2014comparison, wang2014clathrate,zeinelabdein2018critical, lund20144th,khan2017towards, fu2022high}\@. Phase change materials suffer from low thermal conductivity, which hampers their performance \cite{hasnain1998review}\@. Several methods are used to enhance the thermal conductivity of PCMs, such as incorporating metallic fillers, utilizing finned tubes, and adding nanoparticles \cite{abhat1976experimental, abhat1983low, elgafy2005effect}, enhancing the thermal conductivity of liquids can be achieved also by self-propelled particles \cite{el2017self,el2020self} \@.\\

NePCM is a type of phase change material containing particles ranging in size from nanometers to micrometers\@. In its liquid phase, NePCM can be regarded as a colloidal suspension\@. Colloidal suspensions are well-studied materials, with numerous theoretical and experimental investigations aimed at understanding their behaviour \cite{morrison2002colloidal}\@. Colloidal suspensions differ from atomistic solutions in that their thermo-physical properties vary significantly with the volume fraction of the particles \cite{peppin2008experimental}\@.
The physics of colloidal suspensions' solidification (freezing) is still not well understood\@. During the freezing process, particles are significantly rejected from the frozen portion of the suspension due to thermodynamic constraints \cite{dantzig2016solidification}\@. This particle rejection leads to an unstable shape of the solid-liquid interface, enabling the formation of dendrites due to constitutional supercooling, a phenomenon well-documented in many experiments \cite{peppin2006solidification, deville2009metastable, you2018controls, you2016interfacial, you2015situ}\@. Many investigators have employed X-ray tomography and X-ray scattering to gain insights into the shape of the solid-liquid interface and the redistribution of particles during the freezing of colloidal suspensions. These techniques allow visualization of the emergent solid-liquid interface and they map  areas with significant particle concentration \cite{deville2009metastable, spannuth2010structure}\@. Thermosolutal convection is a phenomenon that has yet to be thoroughly investigated for the cases that  phase change occur\@. Thermosolutal convection  emerges  when there are thermal and particle concentration gradients within the system under considerations\@. This type of convection profoundly influences the solidification process of colloidal suspensions, as indicated by experiments conducted by \cite{scotti2017directional, scotti2019effect, delattre2014unidirectional}\@. The experiments that are used for the investigation of the freezing process of the NePCM are primitive compared to one used to investigate colloidal suspensions in the material science community\@. \\

In modelling the freezing process of NePCM, the energy storage community has predominantly employed single-phase models, incorporating the effects of particles primarily into the thermophysical properties\@. For instance, Fan and Khodadadi \cite{fan2012theoretical} utilized a well-known analytical solution for the one-dimensional unidirectional freezing from the bottom, known as the Stefan problem, to study NePCM with varying particle loadings\@. They compared their theoretical predictions with experiments conducted using the same geometry\@.
Their theoretical predictions suggested that NePCM would freeze faster compared to the pure solvent\@. However, their experimental results revealed a non-linear dependence of the expedited freezing of NePCM on the volume fraction of particles, which contradicted their developed theory\@. They attributed this non-linear behaviour to the sedimentation of particles\@.  However,  since the experiment is designed to operate in diffusive conduction conditions\@ Therefore, the layers formed by the highly conductive particles sedimentation should aid in speeding up the freezing process of the NePCM, rather than slowing it down\@. Similar observations to those of Fan and Khodadadi \cite{fan2012theoretical} were made by Murugan et al. \cite{murugan2018thermal}, who reported that NePCM with a loading of 0.3\% carbon nanotubes by volume solidified faster than NePCM with a 0.9\% loading by volume\@. The failure of Fan and Khodadadi's one-dimensional model \cite{fan2012theoretical} to capture and explain the non-linear dependence of expedited freezing of NePCM on the volume fraction of particles in comparison to pure PCM highlights that the mathematical frameworks of single-phase models cannot fully capture the physics involved in the freezing process of NePCM\@. Do to the incompetence of the single phase model to capture the physics that are involved numerical simulations that are using this type of  model are in their majority  erroneous \cite{sheikholeslami2018finite, sheikholeslami2018numerical, sheikholeslami2018solidification, li2019solidification}\@. \\

Following the miss match between the theory, and experiments El Hasadi and Khodadadi\cite{el2013one} developed  a version of the  Stefan problem that includes mass transfer known alternatively as Rubinstein problem suitable to model the solidification process of colloidal suspensions\@. The model predicted the non-linear dependence of the expedited freezing process similar to observations made by Fan and Khodadadi \cite{fan2012theoretical} in their experiments, also the its numerical predictions of the experimental data of Fan and Khodadadi\cite{fan2012theoretical} are better compared to the theoretical model of Fan and Khodadadi\cite{fan2012theoretical}\@. The enhancement of the prediction accuracy of the El Hasadi and Khodadadi \cite{el2013one} model indicates  the importance of adding the mass transfer of particles in the numerical framework that models the phase change process of the NePCM\@. Also, El Hasadi and Khodadadi\cite{el2013one}   showed that the solidification process of the NePCM is governed by  the thermal transport, and  particle  transport due to particle  rejection from the solid-liquid interface, and the parameter that controls the process is the transition segregation coefficient $k_t$\@. $k_t$ is given from the following relation: 
\begin{equation}\label{segregation_trans}
k_t = \frac{\phi_w m_l}{T_c-T_m}
\end{equation}   
Where $\phi_w$, $m_l$, $T_c$, and $T_m$ represent the mass fraction of the particles, the liquidus slope, the cold-side temperature of the system, and the melting temperature of the solvent, respectively\@. The physical interpretation of the transition segregation coefficient is that the solidification process of colloidal suspensions or that of the  NePCM is governed by the removal of heat only if the segregation coefficient $k_0$, defined as the ratio between the mass fraction of the particles in the solid and liquid sides of the interface, is greater than the transition segregation coefficient. However, if the segregation coefficient $k_0$ is lower than the transition coefficient, the solidification process is controlled by the mass transport of the particles. Consequently, any enhancements in thermal conductivity will not expedite the freezing process if $k_o \leq k_t$, explaining the deceleration observed in various experiments on the freezing of NePCM\cite{fan2012theoretical}\@.
El Hasadi and Khodadadi\cite{el2013numerical,el2015numerical} extended the one-dimensional model they developed for the freezing process of colloidal suspensions to `two  dimensional space\@. Additionally, they incorporated the effect of convection into their model. Their findings highlight the essential role played by the size and concentration of the particles in the development of the solid-liquid interface\@. El Hasadi and Khodadadi\cite{el2013numerical,el2015numerical} were the first to predict the dendritic, unstable shape of the solid-liquid interface observed during freezing experiments with colloidal suspensions\cite{deville2009metastable}\@. Furthermore, El Hasadi\cite{el2013numerical,el2015numerical} were also, the first to predict the existence of thermosolutal convection, which was later confirmed by sophisticated experiments conducted by Scotti et al.\cite{scotti2017directional}. The prediction of thermosolutal convection by the model of El Hasadi and Khodadadi\cite{el2013numerical,el2015numerical} underscores the necessity of including mass transfer of particles in the modelling of the freezing process of colloidal suspensions and NePCM\@. Only a few investigations in the literature regarding the freezing process of colloidal suspensions and NePCM include the transport of particles\cite{soni2021nanoadditive, jegatheesan2022model,patil2023effect}\@. \\

For the case of melting the solid form of colloidal suspensions  NePCM, in contrast to the case of freezing colloidal suspensions, where we had a significant number of investigations coming from the physical sciences, we will rely on the research being conducted by the energy storage community\@. \\

There have been numerous experimental investigations exploring the melting process of NePCM using different types of solvents and nanoparticles \cite{chintakrinda2011direct, ho2013experimental, fan2014heat, li2020experimental}. Chintakrinda et al. \cite{chintakrinda2011direct} performed  tests on a NePCM composed of organic PCM and graphene nanofibers. They found that suspending 11\% of graphene suppressed the  convection currents and  did not led to any enhancement in the melting process, despite the addition of thermal conductivity enhancers such as graphene nanofibers. 
Ho and Gao \cite{ho2013experimental} experimentally investigated the performance of n-octadecane and alumina nanoparticles, confirming similar results to Chintakrinda et al. \cite{chintakrinda2011direct}. They observed that increasing the nanoparticle loading significantly degraded the melting performance of the NePCM.
Fan et al. \cite{fan2014heat} and Li et al. \cite{li2020experimental} investigated  the melting process of nano-enhanced phase change materials (NePCM) within a rectangular cavity, heated from the bottom, and vertical sides respectively\@.
 They used cavities with different geometries and varying aspect ratios, examining NePCM samples prepared by incorporating graphene nanoplatelets (GNP) into 1-tetradecanol. To stabilize the NePCM colloidal suspensions, they employed surfactants. The addition of these surfactants usually increase the suspension's viscosity by forming polymeric brushes around the particle surfaces, thereby impeding fluid movement. For instance, Fan et al. \cite{fan2014heat} noted a tenfold increase in NePCM viscosity upon adding 3\% GNP particles compared to that of 1-tetradecanol. This increase in viscosity significantly surpassed what the dilute colloidal hydrodynamic theory predicts, indicating that the addition of the surfactant was the primary reason for the viscosity increase. 
The two previous experimetal investigations' results indicated that an increase in GNP loading led to a decline in both heat storage and heat transfer rates during melting. This effect was attributed to a significant rise in viscosity, which negatively impacted natural convective heat transfer. These findings suggest that utilizing NePCM in a rectangular cavity geometry may not necessarily enhance heat storage rates due to the adverse impact of increased viscosity, which outweighs the thermal conductivity enhancement. Moreover, the  experimental investigations imply that numerical predictions, as we will demonstrate shortly, regarding melting accelerations and heat transfer enhancements reported in the literature by adding nano-paeritles to phase change materilas, may have been overestimated, primarily due to an underestimation of the effects of mass transport of the particles in their models\@. Additionally, a deceleration of the solid-liquid interface during the NePCM melting process was observed in the experiments by \cite{li2020revisiting, li2022synergistic}.\\

Numerical simulations of the melting process of NePCM often rely on the single-phase model, which neglects the transport of particles, including them solely in the transport properties. Xiong et al.'s extensive review \cite{xiong2020nano} on numerical investigations related to the modeling of NePCM's phase change process and Dhaidan and Khodadadi's review \cite{dhaidan2015melting} on experimental investigations into the NePCM melting process both highlight a significant discrepancy between numerical and experimental predictions. They partially attribute this discrepancy to the absence of particle transport in numerical simulation models. Feng et al. \cite{feng2015numerical} utilized the Lattice Boltzmann numerical method to study the melting process of NePCM heated from the bottom side of a rectangular cavity. They reported the classical finding that increasing the volume fraction led to a significant enhancement in the NePCM melting process compared to the pure solvent. However, this enhancement contradicts the findings of many experimental investigations, such as that of Fan et al. \cite{fan2014heat}. The Lattice Boltzmann method, in its current form, does not offer new physical insights into the NePCM melting process due to its inability to account for the mass transport of particles or their interaction with the solid-liquid interface. Practically speaking, both the Lattice Boltzmann method and the single-phase model based on finite volume yield similar physical insights into the NePCM melting process, offering no advantage in using the former. Numerous numerical investigations employ the single-phase model to simulate the NePCM melting process \cite{han2022nanoparticles, zhang2022improving, ghalambaz2019natural, chavan2023melting}. However, most of their results contradict experimental observations. \\

El Hasadi \cite{el2019numerical} was the first to investigate the mass transport of particles during the  melting process of colloidal based NePCM using a similar model as that of  of El Hasadi and Khodadadi  \cite{el2013numerical,el2015numerical} which is based on the  one-fluid mixture model\@. He used a rectangular cavity as a geometry of cohice, the NePCM is melted from the the vertical left side\@. The major findings  were that the solid-liquid interface takes unstable shape, and that the size of the particles plays an essential role in the thermal transport, and into the redistribution of the particles as well. Also, another outocme of the  investigation is the report about the  existence of the thermo-solutal convection currents which  thier intensity depends on the mass fraction of the particles\@. In another investigation, El Hasadi \cite{el2023computational} further investigated the melting process of NePCM within a rectangular cavity heated from two sides, cooled from one side, and insulated on the last side. He observed that an increase in the volume fraction of particles resulted in the deceleration of the NePCM melting process as particle loading increased. For instance, at a particle volume fraction of 10\%, the NePCM's melting process slowed down by approximately 2.2\% compared to the pure solvent, which in this case was water, aligning closely with observations from experiments.\\

In this study, we delve into the influence of thermosolutal convection on the melting process of NePCM within a square cavity geometry, subjected to melting from both the top and bottom sides. We employ a numerical model akin to those employed in \cite{el2013numerical} \cite{el2019numerical}, which incorporates the transport of nanoparticles\@. Distinct values for particle concentration, imposed temperature boundary conditions, and segregation coefficients are implemented to examine the impact of these factors on the melting process\@.

\section{Mathematical Model}

One of our foundational assumptions hinges on treating the NePCM as a binary mixture, a simplification that significantly streamlines the complexity of our mathematical model. By adopting this assumption, we notably disregard direct interaction forces between particles and between particles and the fluid\@. The particle interactions will be preserved in the values of thermo-physical properties\@. The current investigation employs a model built upon the author's prior work and it was  undergone  rigorous testing and validation across various scenarios of NePCM solidification and melting \citep{el2013numerical,el2013one,el2015numerical, el2019numerical,el2013numerical2,el2011numerical,el2012numerical}\@. The basic governing equations are the following:\\ 
\textbf{Continuity:} 
\begin{equation}\label{eq1}
\nabla \cdot \vec{U}=0,
\end{equation} \label{eq2}
\textbf{Momentum}
\begin{equation}
\rho \frac{\partial \vec{U}}{\partial t}+\rho(\vec{U} \cdot \nabla) \vec{U}=-\nabla p+\mu \nabla^{2} \vec{U}-\rho \frac{C_{m}(1-\lambda)^{2}}{\lambda^{3}} \vec{U}+(\rho \beta) g\left(T-T_{\text {ref }}\right) \vec{e}_{y}, 
\end{equation}
\textbf{Energy:}
\begin{equation}\label{eq3}
\rho c_{p} \frac{\partial T}{\partial t}+\rho c_{p} \vec{U} \cdot \nabla T=\nabla \cdot(k \nabla T)-\rho L \frac{\partial \lambda}{\partial t},
\end{equation}
\textbf{Species:}
\begin{equation}\label{eq4}
\frac{\partial \phi_{w}}{\partial t}+\vec{U} \cdot \nabla \phi_{w}=\nabla \cdot\left(D_{B} \nabla \phi_{w}+D_{T} \frac{\nabla T}{T}\right)
\end{equation}
Where $\vec{U}$ represents the velocity vector, $T$ denotes temperature, and $\phi_{w}$ stands for the mass fraction of the particles. The Darcy term within the momentum equations assumes a pivotal role in modulating velocity values during the melting process. This term functions as a damping mechanism, diminishing flow rates in scenarios characterized by high porosity or permeability. Consequently, it prevents instabilities, ensuring a stable and controlled melting process. Within the mushy zone, the local mass fraction is computed using a straightforward mixture formula as follows:
\begin{equation}\label{eq5}
\phi_{w}=\lambda \phi_{w f}+(1-\lambda) \phi_{w s},
\end{equation}
Where $\lambda$ represents the liquid fraction, and the subscripts $f$ and $s$ denote the liquid and solid states, respectively. The momentum equations incorporate Darcy law damping terms, wherein we assign a porosity constant $C_m$ with a value of $10^5 kg/m^3s$\@. Additionally, we disregard particle diffusion within the heat flux vector due to the nano-particles' low diffusion coefficient \citep{buongiorno2006convective}.
\subsection{Initial and boundary conditions}
In our current investigation, we'll focus on a square cavity geometry with thermally insulated left and right sides. For our simulations, we'll explore two variations of this geometry.  In the first configuration, the melting process of the NePCM solid composite will commence from the top side of the cavity. While, in the second configuration, the NePCM solid composite will initiate melting from the bottom side of the cavity. 
Both the bottom and top sides of the cavity will have a constant temperature boundary condition, and the verical sides of the cavity will be isolated thermally. The NePCM solid composite will start melting by elevating the temperature above the melting temperature for the given particle concentration\@. We'll  use various nano-particle concentrations, with a particle size of $d_p = 2nm$. The geometry under-consideration is shown in Figure \ref{geomtery}\@.
\subsection{Effective thermo-physical properties}
The NePCM mixture that we will use  is suspension of water, and copper nano-particles\@.  
We'll make the assumption that the suspension in its liquid form comprises hard-sphere particles, thereby preventing particle clustering. Additionally, we'll consider the properties of the pure water in solid and liquid phases as constant for simplification purposes. However, these properties vary with the concentration of the particles, a characteristic often observed in colloidal particles. \\

To derive the density, heat capacity, and a portion of the Boussinesq term, we'll utilize the following equations.
\begin{equation} \label{eq6}
\begin{aligned} 
&\rho=(1-\phi) \rho_{f}+\phi \rho_{p} \\
&\rho c_{p}=(1-\phi)\left(\rho c_{p}\right)_{f}+\phi\left(\rho c_{p}\right)_{p}, \\
&\rho \beta=(1-\phi)(\rho \beta)_{f}+\phi(\rho \beta)_{p} .
\end{aligned}
\end{equation}
Where $\phi$, $\rho$, and $\beta$ represent the volume fraction of the particles, the density, and the thermal expansion coefficient, respectively. The subscripts $f$ and $p$ denote the liquid (solid) and particle phases, respectively. The relationship between the volume fraction of the particles and their mass fraction is as follows:
\begin{equation}\label{eq7}
\phi=\frac{\rho_{f} \phi_{w}}{\rho_{f} \phi_{w}+\rho_{p}\left(1-\phi_{w}\right)}
\end{equation}
The thermal conductivity of the NePCM is calculated as the following: 
\begin{equation}\label{eq8}
k=k_{1}+k_{2}
\end{equation}
Where $k_1$ is the Maxwell contribution to the thermal conductivity: 
\begin{equation} \label{eq9}
k_{1}=k_{f} \frac{k_{p}+2 k_{f}-2 \phi\left(k_{f}-k_{p}\right)}{k_{p}+2 k_{f}+\phi\left(k_{f}-k_{p}\right)}
\end{equation}
and $k_2$ is the contribution due to the dispersion:
 \begin{equation}\label{eq10}
 k_{2}=C_{k}\left(\rho c_{p}\right)|\vec{U}| \phi d_{p}
 \end{equation}
The redistribution of particles will notably impact the mass diffusivity of the particles. To account for this effect, we will define a compressibility factor as a function of the volume fraction, following  Peppin et al. \cite{peppin2006solidification}\@. The constant $C_k$ is obtained from Wako and Kaguei \cite{wakao1982heat}\@. The compressibility equation is expressed as follows:

\begin{equation} \label{eq11}
z(\phi)=\frac{1+\left(4-\frac{1}{0.64}\right) \phi+\left(10-\frac{4}{0.64}\right) \phi^{2}+\left(18-\frac{10}{0.64}\right) \phi^{3}}{1-\frac{\phi}{0.64}}
\end{equation}
The mass diffusivity of the particles can be calculated from the following equation: 
\begin{equation} \label{eq12}
D_{B}=D_{o}(1-\phi)^{6}\left(\frac{d(\phi z)}{d \phi}\right)
\end{equation}
Where $D_0$ is the Einstein diffusivity:
\begin{equation}\label{eq13}
D_{0}=\frac{T k_{B}}{3 \pi d_{p} \mu}
\end{equation}
The theromophoretic diffusivity is evaluated from the following relation:
\begin{equation}\label{eq14}
D_{T}=\beta_{k}\left(\frac{\mu}{\rho}\right) \phi
\end{equation}
Where $\beta_k$ is a non-dimensional parameter that is a function of the thermal conductivities described by \cite{buongiorno2006convective}, the variation of the latent heat of fusion of the NePCM is given as the following:
\begin{equation}\label{eq15}
\rho L=(1-\phi)(\rho L)_{f}
\end{equation} 
To incorporate the change  of the liquidus and solidus temperatures of the NePCM with the  volume fraction of the particles, we'll employ the following straightforward linear phase diagram:
\begin{equation}\label{eq16}
\begin{aligned}
&T_{\text {Liq }}=T_{m}-m_{l} \phi_{w} \\
&T_{\text {Sol }}=T_{m}-\frac{m_{l}}{k_{0}} \phi_{w}
\end{aligned}
\end{equation}
Where $T_{\text{Liq}}$ and $T_{\text{Sol}}$ represent the solidus and liquidus temperatures of the NePCM suspension, respectively. $k_0$ denotes the segregation coefficient, governing the ratio of the mass fraction of the particles on the solid and liquid sides of the interface. For our simulations, we'll set the {segregation coefficient to a value of 0.1}. We'll utilize the following equation to calculate the melting temperature:

\begin{equation} \label{eq17}
m_{l}=\frac{k_{B} T_{m}^{2}}{v_{p} \rho L_{f}}
\end{equation}
The relation that relates the liquid fraction $\lambda$ with the $T_{\text{Liq}}$, and $T_\text{Sol}$ temperatures is the following:
\begin{equation}
\begin{aligned}
& \lambda = 0 &\text{for}&&\ T < T_\text{Sol}\\
& \lambda =\frac{T-T_\text{Sol}}{T_{\text{Liq}}-T_\text{Sol}} &\text{for}&&\ T_\text{Sol}<T<T_{\text{Liq}}\\
& \lambda = 1  &\text{for} &&\ T>T_{\text{Liq}}\\
\end{aligned}
\end{equation}

The thermophysical properties of water and copper particles are listed in Table \ref{Tab1}. Additionally, Table \ref{Tab2} presents the properties of the NePCM for a mass fraction of 10\%. From Table \ref{Tab2}, it is evident that even at the maximum particle mass fraction that we will use, both the thermal conductivity and viscosity do not exhibit significant enhancements compared to pure water.
\begin{table} [H]
\begin{tabular}{lll}
\hline & Water & Copper nanoparticles \\
\hline Density & $997.1 \mathrm{~kg} / \mathrm{m}^{3}$ & $8954 \mathrm{~kg} / \mathrm{m}^{3}$ \\
Viscosity & $8.9 \times 10^{-4} \mathrm{~Pa} \mathrm{~s}$ & $-$ \\
Specific heat & $4179 \mathrm{~J} / \mathrm{kg} \mathrm{K}$ & $383 \mathrm{~J} / \mathrm{kg} \mathrm{K}$ \\
Thermal conductivity & $0.6 \mathrm{~W} / \mathrm{m} \mathrm{K}$ & $400 \mathrm{~W} / \mathrm{m} \mathrm{K}$ \\
Thermal expansion coefficient  & $2.1 \times 10^{-4} \mathrm{~K}^{-1}$ & $1.67 \times 10^{-5} \mathrm{~K}^{-1}$ \\
Heat of fusion & $3.35 \times 10^{5} \mathrm{~J} / \mathrm{kg}$ & $-$ \\
\hline
\end{tabular}
\caption{Thermophysical and transport properties for the solvent and the nanoparticles}
\label{Tab1} 
\end{table}
\begin{table} [H]
\begin{center}
\begin{tabu} to 1.0\textwidth { | X[c] | X[c] | }
 \hline
 Transport Property & $\phi_w$ = 10\% ($\phi$=\eval* {$\frac{\rho_f\phi_w*100}{\rho_f\phi_w+\rho_p(1-\phi_w)}$}
 [\rho_f = 997.1,\phi_w = 0.1,\rho_p = 8954][x2]\%) \\ 
 \hline
Thermal Conductivity   & 6.23$\times 10^{-1}$ $ \mathrm{~W} / \mathrm{m} \mathrm{K}$  \\
 Density  & 1.10$\times 10^{3}$ $\mathrm{~kg} / \mathrm{m}^{3}$    \\
 Dynamic viscosity   & 9.18 $\times 10^{-4}$ $\mathrm{~Pa} \mathrm{~s}$    \\
 Specific heat  & 3.8$\times 10^3$ $\mathrm{~J} / \mathrm{kg} \mathrm{K}$ \\
 Thermal expansion coefficient  & \eval*{$10^{4}\frac{(1-\phi)(\rho_f\beta_f)+\phi(\rho_p\beta_p)}{\rho_m}$}[\phi=0.0122,\rho_f = 997.1,\beta_f = 0.00021,\rho_p = 8954, \beta_p = 0.0000167,\rho_m = 1100][x2] $\times 10^{-4}$ $\mathrm{~K}^{-1}$ \\
\hline

\end{tabu}
\end{center}
\caption{The values of the transport properties of the NePCM at different nano-particle volume fractions.}
\label{Tab2}
\end{table}
\section{Results}

The results will include cases of melting the NePCM from the top, and bottom sides receptively\@. 
\subsection{Melting from the top side}
\begin{figure}[p!]
\centering
\begin{subfigure}{.5\textwidth}
 \centering
 \caption{t= 8 sec}
  \includegraphics[width=.8\linewidth, trim =2.0cm 2.0cm 2.0cm 2.0cm,clip ]{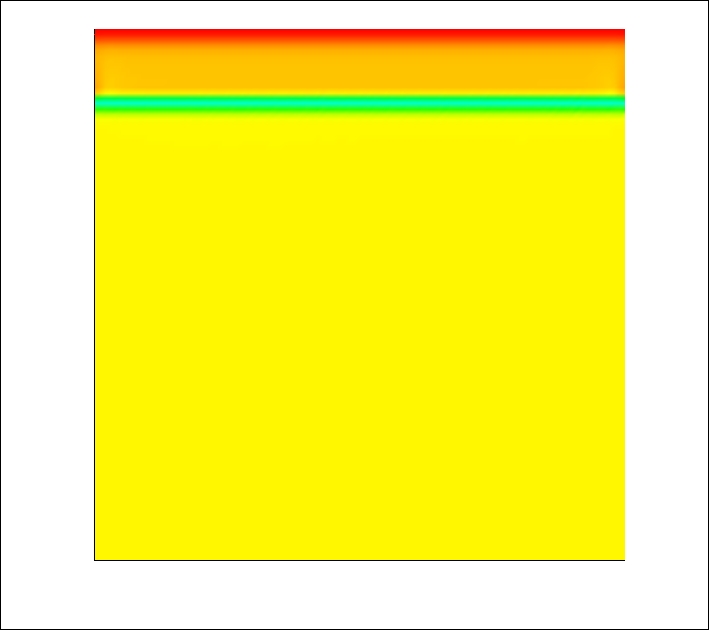}
\end{subfigure}%
\begin{subfigure}{.5\textwidth}
  \centering
  \caption{t = 50 sec}
  \includegraphics[width=.8\linewidth,trim = 2.0cm 2.0cm 2.0cm 2.0cm,clip]{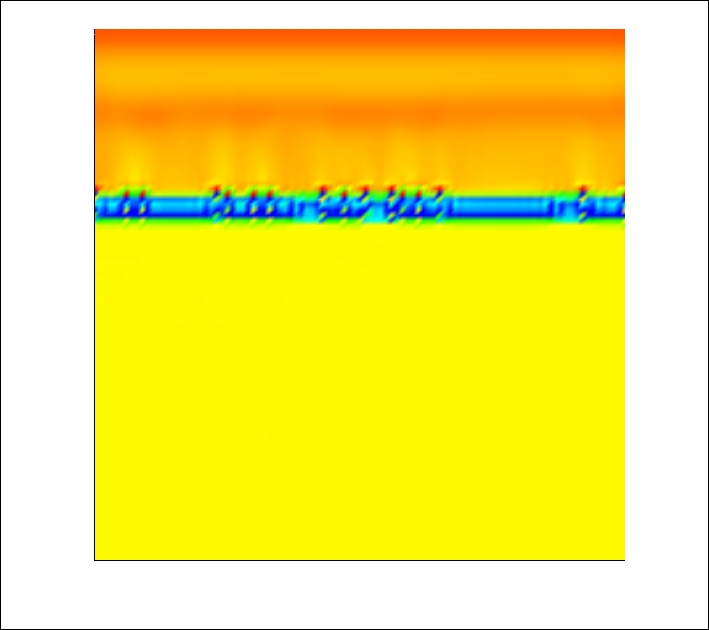}
\end{subfigure}
\bigskip
\begin{subfigure}{.5\textwidth}
  \centering
   \caption{t = 140 sec}
  \includegraphics[width=.8\linewidth,trim = 2.0cm 2.0cm 2.0cm 2.0cm,clip]{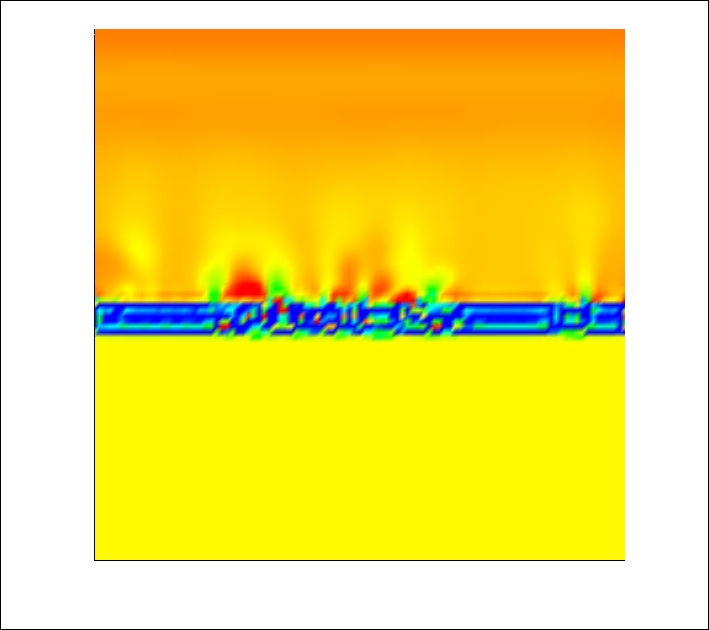}
\end{subfigure}%
\begin{subfigure}{.5\textwidth}
  \centering
  \caption{t = 200 sec}
  \includegraphics[width=.8\linewidth,trim = 2.0cm 2.0cm 2.0cm 2.0cm,clip]{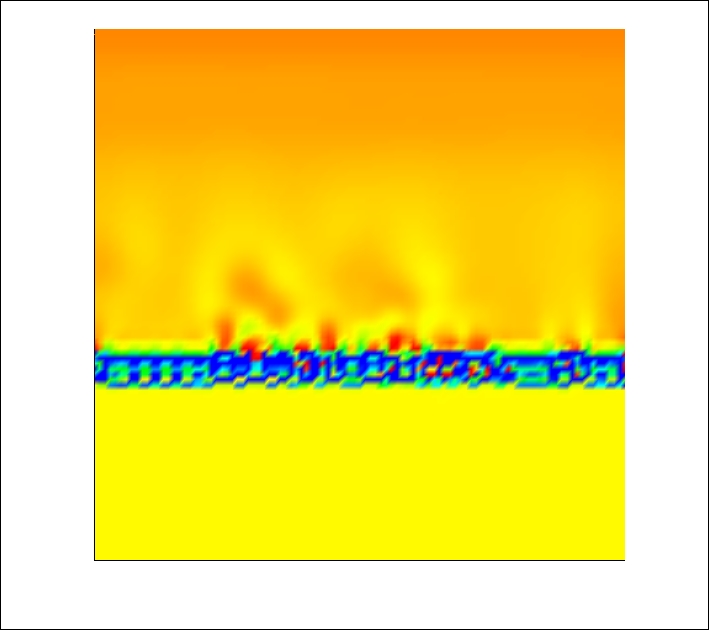}
\end{subfigure}
\bigskip
\begin{subfigure}{.5\textwidth}
  \centering
   \includegraphics[width= 1.2\linewidth,trim = 0.1cm 2.0cm 1.0cm 2.0cm,clip]{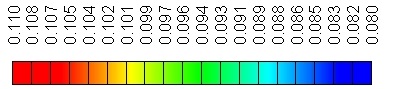}
\end{subfigure}
\caption{The evolution of concentration profiles of nano-particles for the case of melting from the top for  $T_h$ = 290 K, $T_c$ = 270 K, and $\phi_w$ = 10\%.}
\label{Figcontopm}
\end{figure}

\begin{figure}[p!]
\centering
\begin{subfigure}{.5\textwidth}
 \centering
 \caption{t= 8 sec}
  \includegraphics[width=.8\linewidth, trim =2.0cm 2.0cm 2.0cm 1.0cm,clip ]{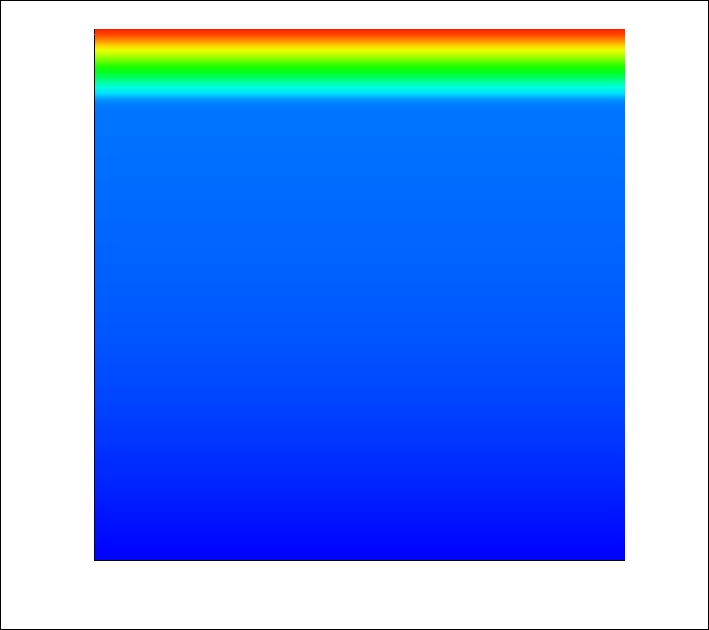}
\end{subfigure}%
\begin{subfigure}{.5\textwidth}
  \centering
  \caption{t = 50 sec}
  \includegraphics[width=.8\linewidth,trim = 2.0cm 2.0cm 2.0cm 2.0cm,clip]{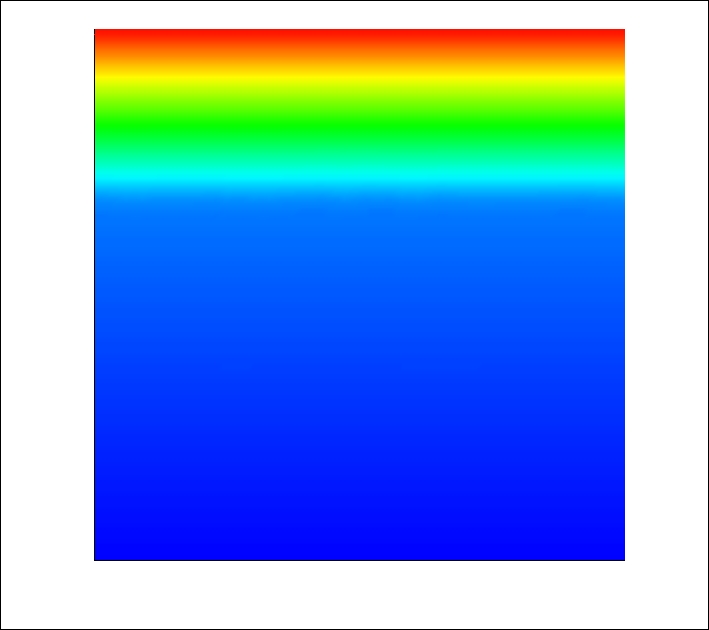}
\end{subfigure}
\bigskip
\begin{subfigure}{.5\textwidth}
  \centering
   \caption{t = 140 sec}
  \includegraphics[width=.8\linewidth,trim = 2.0cm 2.0cm 2.0cm 2.0cm,clip]{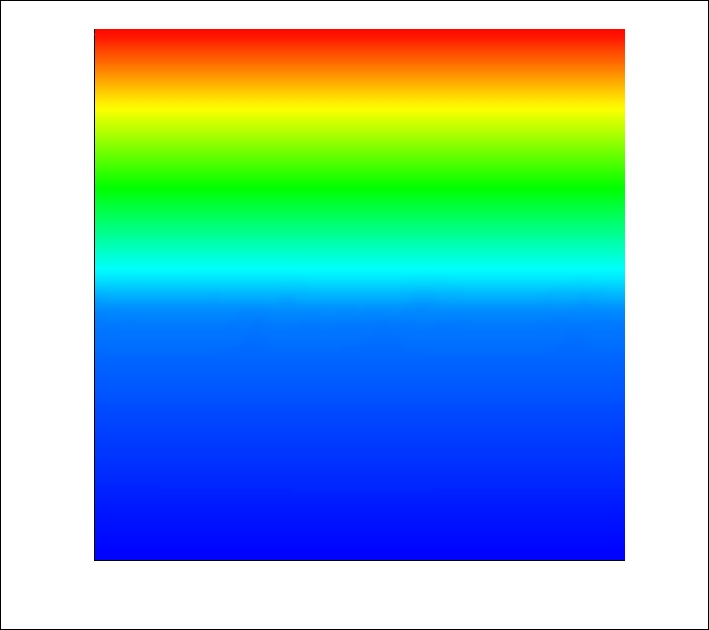}
\end{subfigure}%
\begin{subfigure}{.5\textwidth}
  \centering
  \caption{t = 200 sec}
  \includegraphics[width=.8\linewidth,trim = 2.0cm 2.0cm 2.0cm 2.0cm,clip]{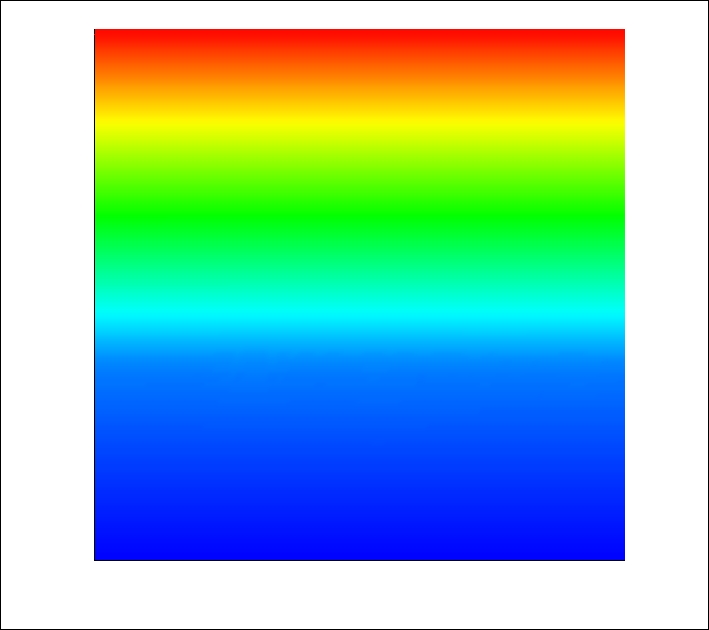}
\end{subfigure}
\bigskip
\begin{subfigure}{.5\textwidth}
  \centering
  \includegraphics[width= 1.2\linewidth,trim = 0cm 0cm 0cm 0cm,clip]{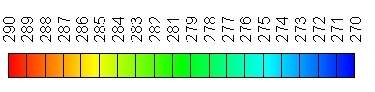}
\end{subfigure}
\caption{The evolution of temperature  profiles  for the case of melting from the top for  $T_h$ = 290 K, $T_c$ = 270 K, and $\phi_w$ = 10\%.}
\label{Figtempetopm}
\end{figure}

In this subsection  will consider a single case with $Ra$ = \eval*{$10^{-4}\frac{\rho g \beta(Th - Tc)H^3}{\mu\frac{k}{\rho cp}}$}[\rho = 1.1\times 10^{3}, g = 9.8,\beta = 1.9\times 10^{-4}, Th = 290,Tc = 270, H = 0.005, \mu = 9.18\times 10^{-4}, k = 6.23\times 10^{-1}, cp = 3.8\times 10^{3}][x2] $\times 10^4$, and specifically will impose the hot side on the top side of the cavity, with $T_h$ = 290 K, while the bottom side is kept at $T_c$ = 270 K,  the initial mass concentration $\phi_w$ is equal to 10\%, and $k_o$ = 0.1\@.  This particular configuration suppresses thermal convection because the temperature field is stably stratified, and thermal conduction will transport the heat\@. The solid-liquid interface develops a stable shape, which is a  particular property of melting controlled by conduction\@. At later times of the melting process, the mushy zone is extended, and we can observe small pockets of an unstable interface, this is a result of the particle rejection from the solid phase to the liquid phase as shown in Figure \ref{Figliquidtopm}\@. The melted part of the NePCM contains a higher concentration of particles, as shown in Figure \ref{Figcontopm}, consistent with experimental observations \cite{dhaidan2017nanostructures}, which   report particle sedimentation during the melting process of the NePCM\@. The high concentration nano-particle pockets are concentrated in the solid interface area, and they remain attached to the interface as the melting process proceeds\@. Since there is no convection mechanism to redistribute the particles away from the solid-liquid interface\@. The temperature contours reflect the domination of  conduction as the main heat transfer mechanism, in which the heat is diffused from the high-temperature top side to the cold bottom side, as shown in Figure \ref{Figtempetopm}.\\

\begin{figure}[p!]
\begin{center}
\includegraphics [scale= 0.8, trim = 100 30 0 0,clip]{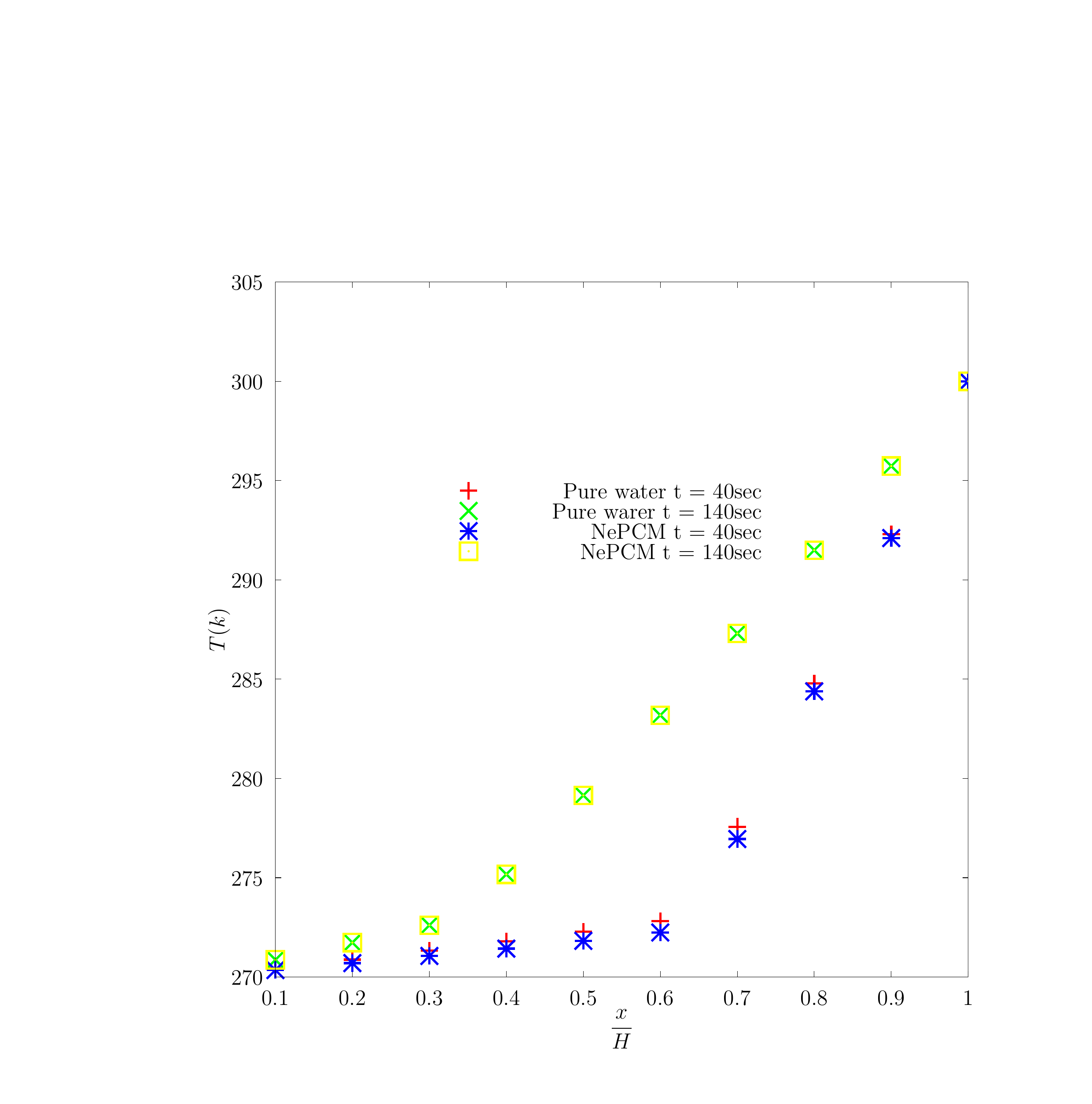}
\end{center}
\caption{The variation of the temperature along a vertical line that crosses the bottom, and top sides of the cavity at $\dfrac{x}{H}$ = 0.5, for the case of for the case of melting from the top for  $T_h$ = 290 K, $T_c$ = 270 K, and $\phi_w$ = 10\%. }
\label{Figcentertoptemp}
\end{figure}

We will compare the thermal performance of the NePCM  with $\phi_w$ =  10\% with that of pure water, by comparing the temperature fields, and the average liquid fraction\@. We will evaluate the temperature profile along a vertical line that crosses the bottom, and top sides of the cavity at the center $\dfrac{x}{H}$, where $H$ is the length of the cavity, and it is equal in our case to 0.005 m\@. The results of the compression is shown in Figure \ref{Figcentertoptemp},  the two temperature profiles  of the NePCM, and that  of  pure water are following each other closely, which shows that adding the particles did not enhance sufficiently the thermal properties especially that of thermal conductivity, that is the main thermal transport property that controls the transport during the conduction process\@. If we compare the values of the thermal conductivities of the pure water, and the NePCM with $\phi_w$ = 10\% from the Table \ref{Tab1}, and Table  \ref{Tab2} we will find that the thermal conductivity only increased about 4.0\%, which is consistent with the effective medium theory, and the findings of the world wide benchmark  exercise for measuring the thermal conductivity of nanofluids \cite{buongiorno2009benchmark}\@.\\

\begin{figure}[p!]
\begin{center}
\includegraphics [scale= 0.8, trim = 100 30 0 0,clip]{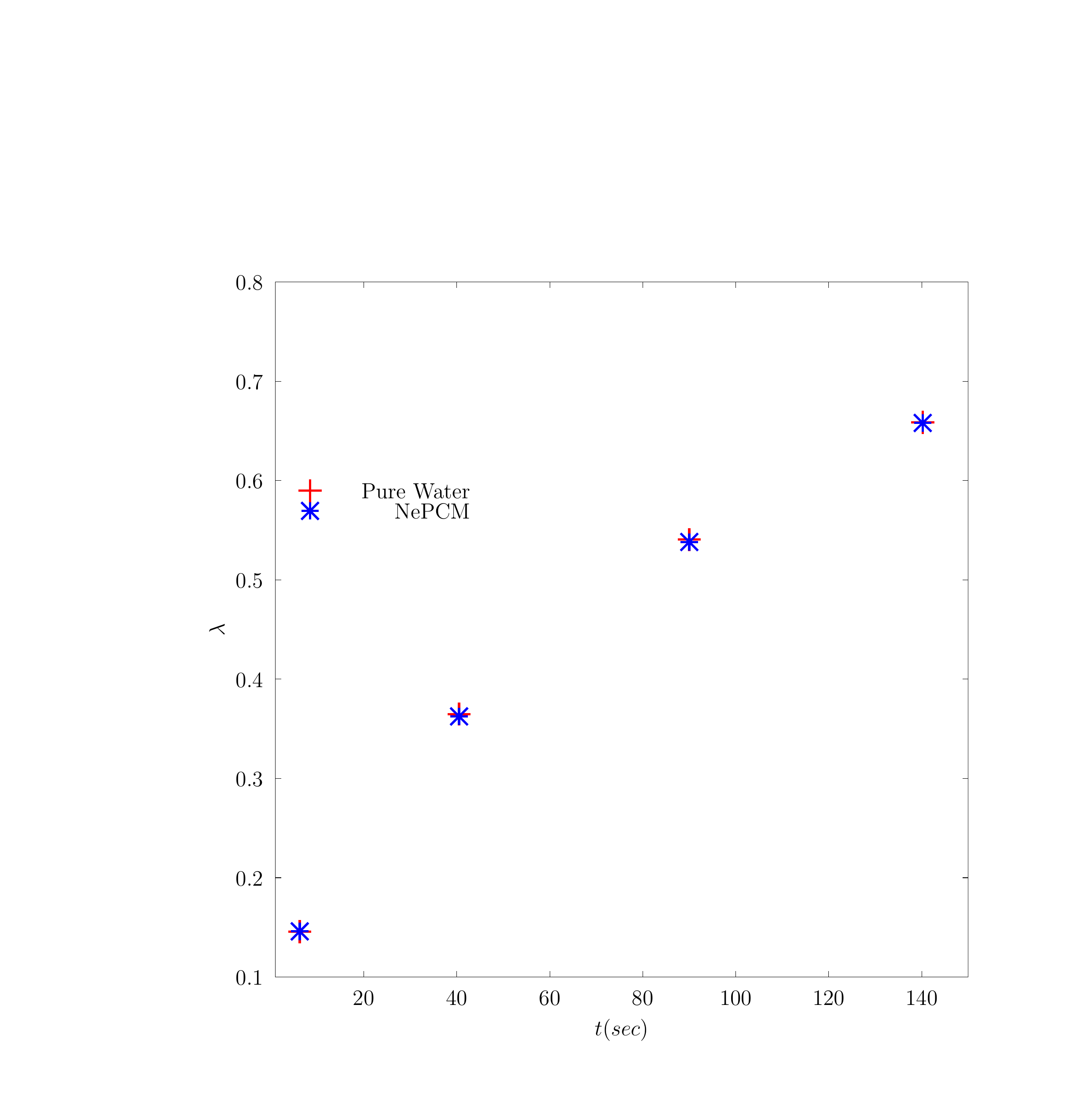}
\end{center}
\caption{Comparison  between the variation of the average liquid fraction for different time intenseness, for the case of pure water, and NePCM $\phi_w$ = 10\%, for  melting from the top with $T_h$ = 290 K, $T_c$ = 270 K. }
\label{Figtimeaveliq}
\end{figure}

 We used low concentration of particles for our simulations, for the two following reasons\@. The first reason is that it will reduce  the computational time needed to perform the simulations\@.While the second reason is that we wanted to follow the recommendations made by different experimental investigations that recommend the  use of suspensions    that contain particle loadings of about 1\% to 5\% by volume,  in order to for the suspensions to stay in stable condition \cite{buongiorno2009benchmark}\@. The volume concentration of particles that we used is 1.22 \% \@. The comparison  of the average liquid fraction for different time instances is shown in Figure \ref{Figtimeaveliq}, for the NePCM, and pure water, the liquid fraction evolution for the two cases is identical indicating of no enhancement in the melting rate\@. Investigations in the literature show enhancement in the melting rate, only when the they use concentrations of particles that are higher than   5\%, for example they used concentration of particles $\phi$ equal to 5\%, and 10\% in their simulations to show enhancement of the performance of the NePCM compared to pure water \cite{feng2015numerical}, suspensions with such high concentration of particles, it will be very difficult to kept in stable condition, due to the exsistance of many types of interparticle, fluid - particles interactions that they will intiate the formation of clusters \cite{mewis2012colloidal}\@.  
\subsection{Melting form the bottom side}
\begin{figure}[p!]
\centering
\begin{subfigure}{.5\textwidth}
 \centering
 \caption{t= 8 sec}
  \includegraphics[width=.8\linewidth, trim =2.0cm 2.0cm 2.0cm 1.0cm,clip ]{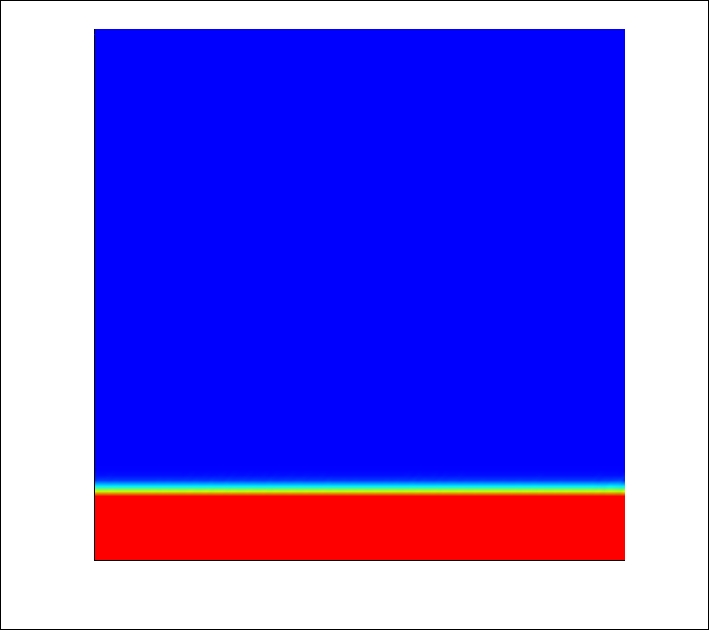}
\end{subfigure}%
\begin{subfigure}{.5\textwidth}
  \centering
  \caption{t = 52 sec}
  \includegraphics[width=.8\linewidth,trim = 2.0cm 2.0cm 2.0cm 2.0cm,clip]{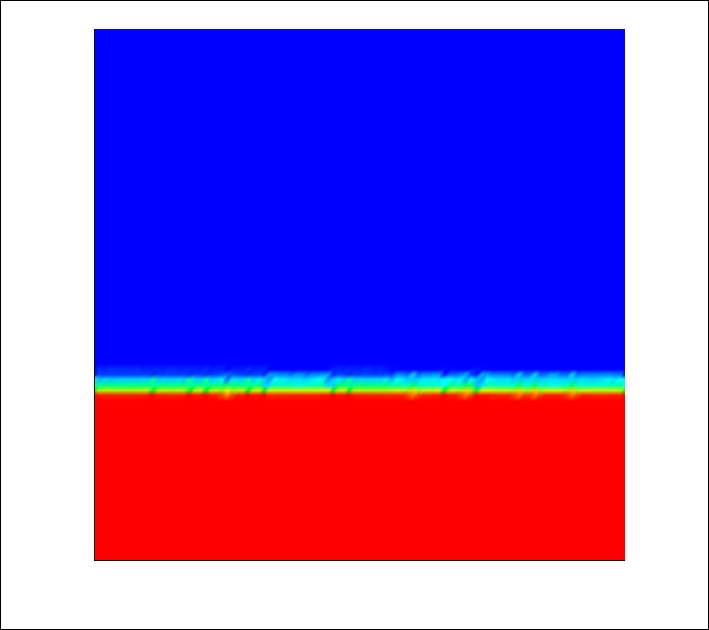}
\end{subfigure}
\bigskip
\begin{subfigure}{.5\textwidth}
  \centering
   \caption{t = 127 sec}
  \includegraphics[width=.8\linewidth,trim = 2.0cm 2.0cm 2.0cm 2.0cm,clip]{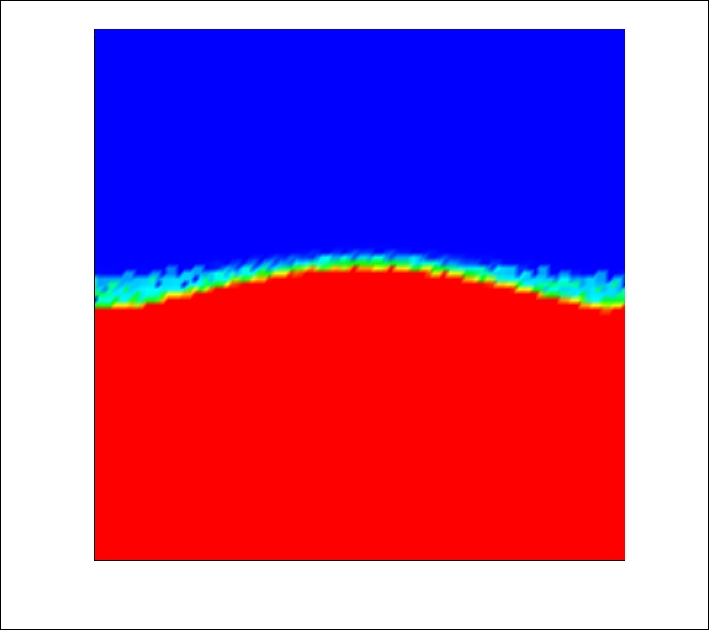}
\end{subfigure}%
\begin{subfigure}{.5\textwidth}
  \centering
  \caption{t = 143 sec}
  \includegraphics[width=.8\linewidth,trim = 2.0cm 2.0cm 2.0cm 2.0cm,clip]{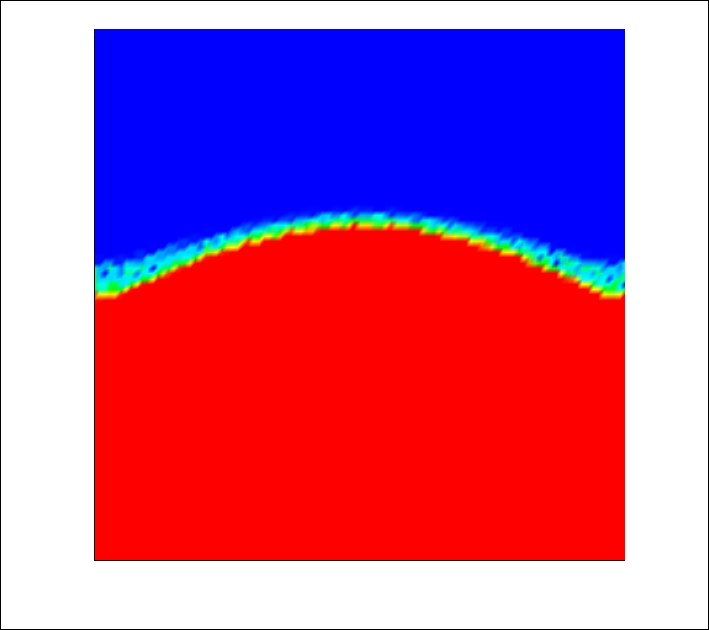}
\end{subfigure}
\bigskip
\begin{subfigure}{.5\textwidth}
  \centering
  \includegraphics[width= 1.2\linewidth,trim = 0cm 0cm 0cm 0cm,clip]{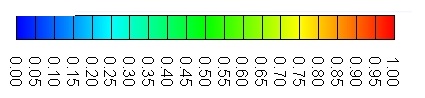}
\end{subfigure}
\caption{The evolution of liquid fraction profiles for the case of melting from the top for  $T_h$ = 290 K, $T_c$ = 270 K, and $\phi_w$ = 10\%.}
\label{Figliquidbottom}
\end{figure}

Now will consider the case of melting from the bottom side, the thermal Rayleigh number equal to \eval*{$10^{-4}\frac{\rho g \beta(Th - Tc)H^3}{\mu\frac{k}{\rho cp}}$}[\rho = 1.1\times 10^{3}, g = 9.8,\beta = 2.1\times 10^{-4}, Th = 290,Tc = 270, H = 0.005, \mu =8.9\times 10^{-4}, k =
6.0\times 10^{-1}, cp = 4.179\times 10^{3}][x2] $\times 10^4$, for the case of pure water, and $Ra$  =  \eval*{$10^{-4}\frac{\rho g \beta(Th - Tc)H^3}{\mu\frac{k}{\rho cp}}$}[\rho = 1.1\times 10^{3}, g = 9.8,\beta = 1.9\times 10^{-4}, Th = 290,Tc = 270, H = 0.005, \mu = 9.18\times 10^{-4}, k = 6.23\times 10^{-1}, cp = 3.8\times 10^{3}][x2] $\times 10^4$\@ for the NePCM with $\phi_w$  = 10\% \@. The value of the $Ra$ for the case of the pure water is higher than that of the NePCM, this is due to the slight increase in the viscosity with volume fraction of the particles, and  combined  with the very marginal increase in the thermal conductivity of NePCM\@.  The hot side is the bottom side of the cavity with $T_h$ = 290 K, while the cold side is the top wall of the cavity, and its temperature is set to ($T_c$) equal to 270 K\@. In this specific configuration, thermal convection will form due to the temperature and density differences\@. The evolution of the melting process for NePCM with a mass fraction of particles ($\phi_w$) = 10\% is shown in Figure \ref{Figliquidbottom}\@. At times $t$ = 8 and 52 seconds, heat is transported through conduction\@. Since the morphology of the solid-liquid interface is flat (planar) in shape, there is no indication that convection has started to form. As we proceed further in time, at $t$ = 127 and 143 seconds, convection begins to form, and the shape of the solid-liquid interface changes from flat to curved due to the effects of natural convection\@.\\

\begin{figure}[p!]
\centering
\begin{subfigure}{.5\textwidth}
 \centering
 \caption{t= 8 sec}
  \includegraphics[width=.8\linewidth, trim =2.0cm 2.0cm 2.0cm 1.0cm,clip ]{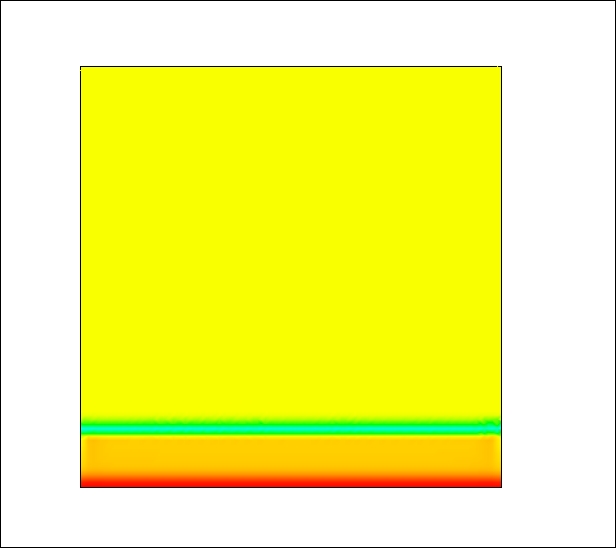}
\end{subfigure}%
\begin{subfigure}{.5\textwidth}
  \centering
  \caption{t = 52 sec}
  \includegraphics[width=.8\linewidth,trim = 2.0cm 2.0cm 2.0cm 2.0cm,clip]{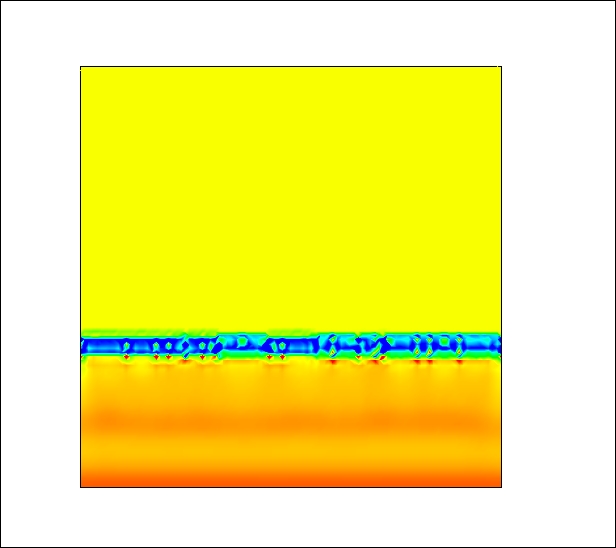}
\end{subfigure}
\bigskip
\begin{subfigure}{.5\textwidth}
  \centering
   \caption{t = 127 sec}
  \includegraphics[width=.8\linewidth,trim = 2.0cm 2.0cm 2.0cm 2.0cm,clip]{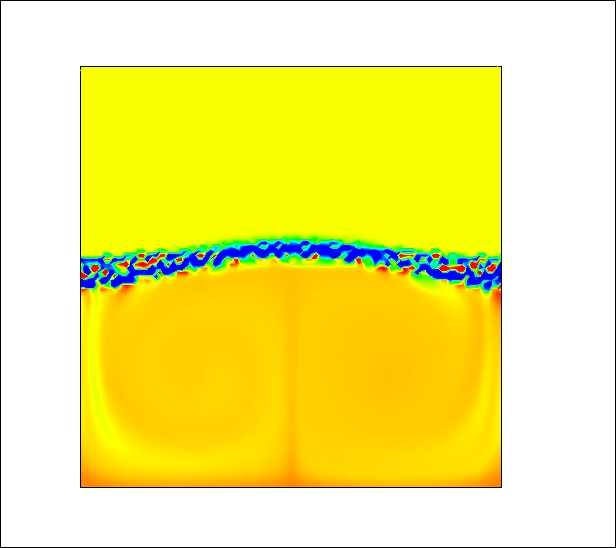}
\end{subfigure}%
\begin{subfigure}{.5\textwidth}
  \centering
  \caption{t = 143 sec}
  \includegraphics[width=.8\linewidth,trim = 2.0cm 2.0cm 2.0cm 2.0cm,clip]{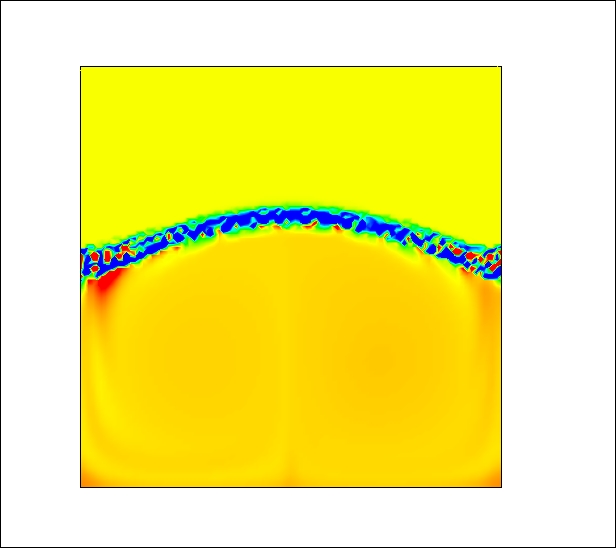}
\end{subfigure}
\bigskip
\begin{subfigure}{.5\textwidth}
  \centering
  \includegraphics[width= 1.2\linewidth,trim = 0cm 0cm 0cm 0cm,clip]{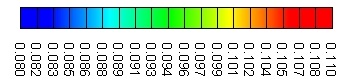}
\end{subfigure}
\caption{The evolution of concentration profiles of nano-particles for the case of melting from the bottom for  $T_h$ = 290 K, $T_c$ = 270 K, and $\phi_w$ = 10\%.}
\label{Figconbottom}
\end{figure}

The accompanying concentration profiles are shown in Figure \ref{Figconbottom}\@. During the early stages of the melting process, layers with different concentrations of particles are formed, with a zone of high particle concentration forming above the hot bottom wall. The concentration of particles is higher in the melting part of the NePCM due to the rejection of the particles from the solid phase during the phase change process\@. At $t$ = 125 and 143 seconds, we observe a significant redistribution of the particles due to the formation of convection cells\@. Convection cells facilitate the dissipation of high particle concentration layers that form due to particle diffusion during the initial stages of melting\@. With the help of convection cells, the particles are more evenly distributed throughout the melting part of the NePCM, and only small high-concentration pockets exist on the left and vertical at the vicinity of the solid-liquid interface\@.\\

\begin{figure}[p!]
\centering
\begin{subfigure}{.5\textwidth}
 \centering
 \caption{t= 8 sec}
  \includegraphics[width=.8\linewidth, trim =2.0cm 2.0cm 2.0cm 1.0cm,clip ]{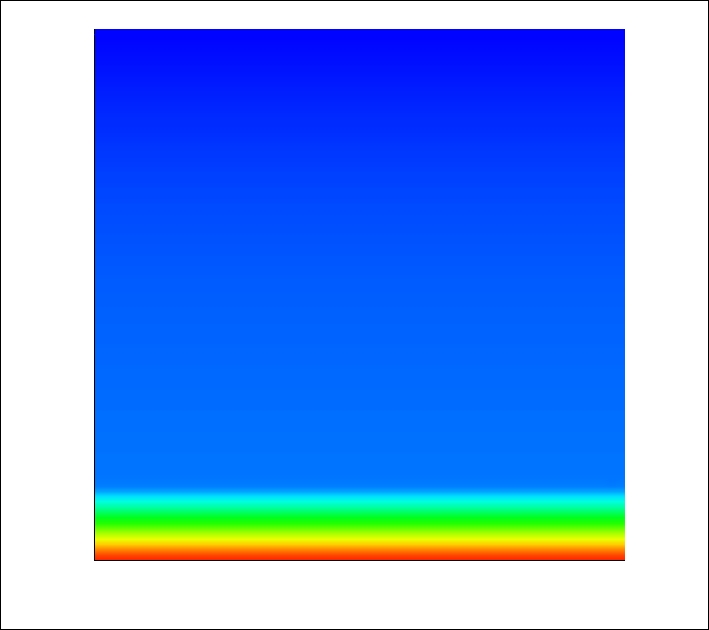}
\end{subfigure}%
\begin{subfigure}{.5\textwidth}
  \centering
  \caption{t = 52 sec}
  \includegraphics[width=.8\linewidth,trim = 2.0cm 2.0cm 2.0cm 2.0cm,clip]{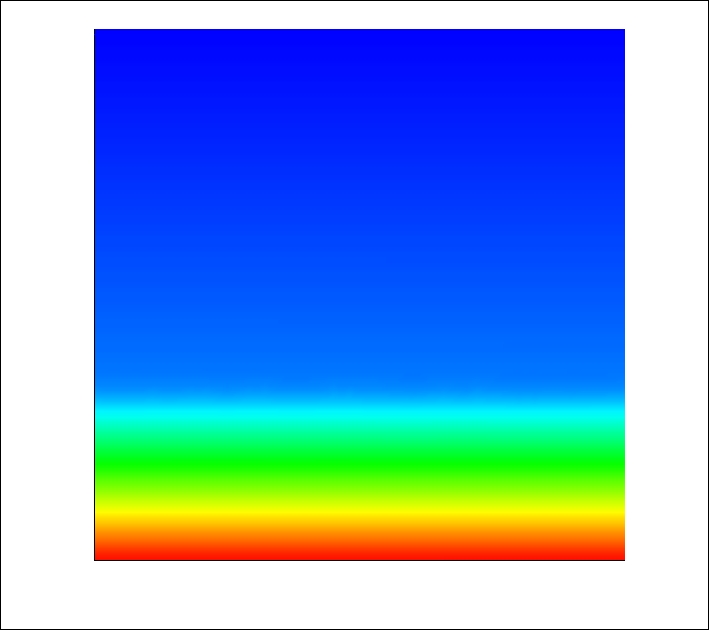}
\end{subfigure}
\bigskip
\begin{subfigure}{.5\textwidth}
  \centering
   \caption{t = 127 sec}
  \includegraphics[width=.8\linewidth,trim = 2.0cm 2.0cm 2.0cm 2.0cm,clip]{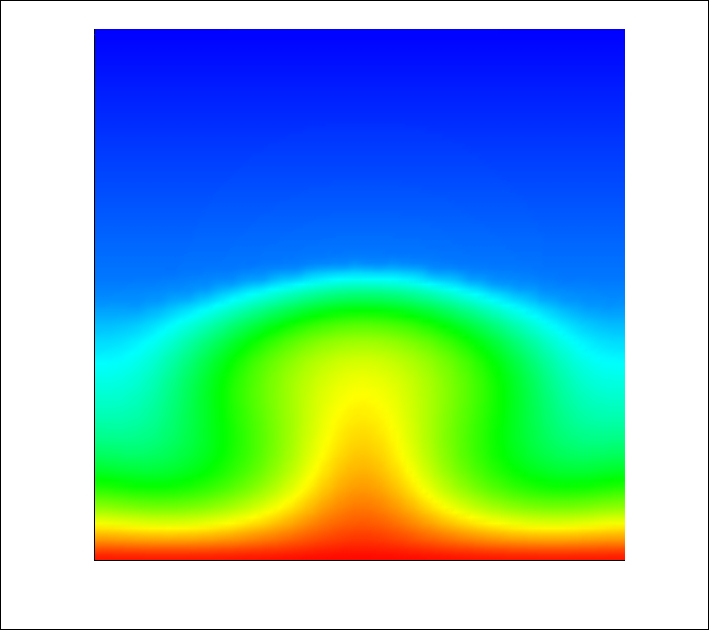}
\end{subfigure}%
\begin{subfigure}{.5\textwidth}
  \centering
  \caption{t = 143 sec}
  \includegraphics[width=.8\linewidth,trim = 2.0cm 2.0cm 2.0cm 2.0cm,clip]{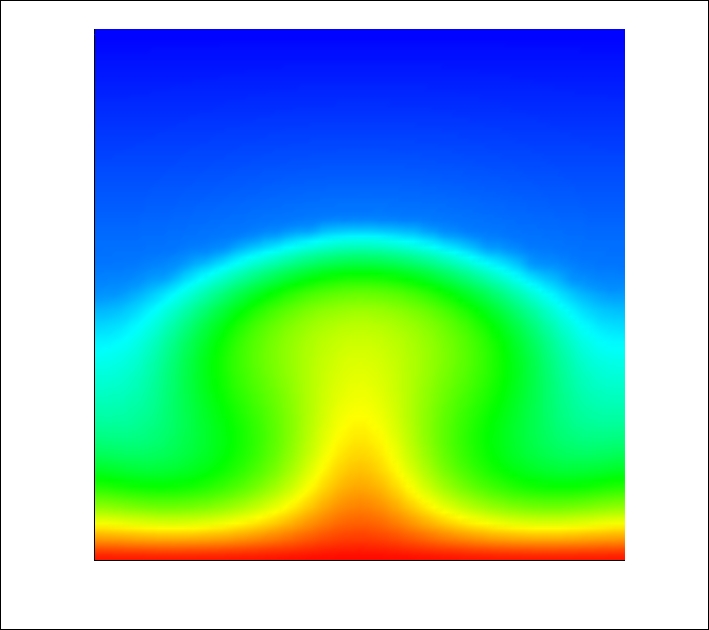}
\end{subfigure}
\bigskip
\begin{subfigure}{.5\textwidth}
  \centering
  \includegraphics[width= 1.2\linewidth,trim = 0cm 0cm 0cm 0cm,clip]{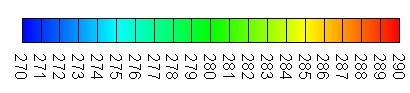}
\end{subfigure}
\caption{The evolution of temperature  profiles  for the case of melting from the bottom for  $T_h$ = 290 K, $T_c$ = 270 K, and $\phi_w$ = 10\%.}
\label{Figtempebottom}
\end{figure}

At times $t$ = 8 and 52 seconds, the temperature is diffusing from the hot bottom side\@. At later times, however, the hot temperature at the bottom wall's centre rises due to convection, bouncing the convection current to warm layers of water away from the bottom wall\@. The temperature profiles for the case of bottom NePCM melting are shown in Figure \ref{Figtempebottom}\@.\\

\begin{figure}[p!]
\centering
\begin{subfigure}{.5\textwidth}
 \centering
 \caption{t= 127 sec}
  \includegraphics[width=.8\linewidth, trim =2.0cm 2.0cm 2.0cm 2.0cm,clip ]{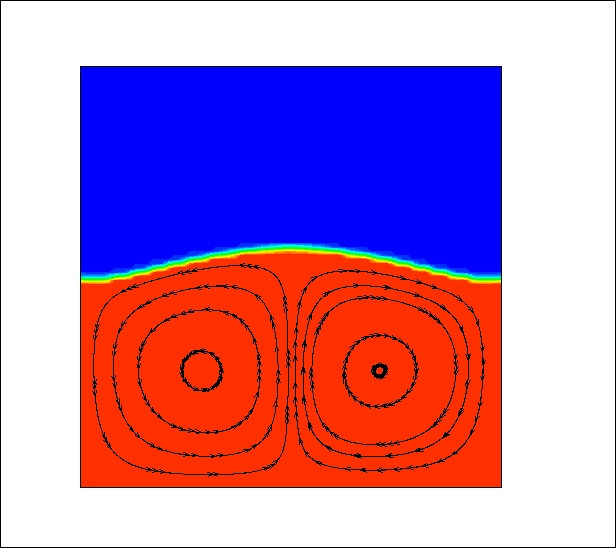}
\end{subfigure}%
\begin{subfigure}{.5\textwidth}
  \centering
  \caption{t = 143 sec}
  \includegraphics[width=.8\linewidth,trim = 2.0cm 2.0cm 2.0cm 2.0cm,clip]{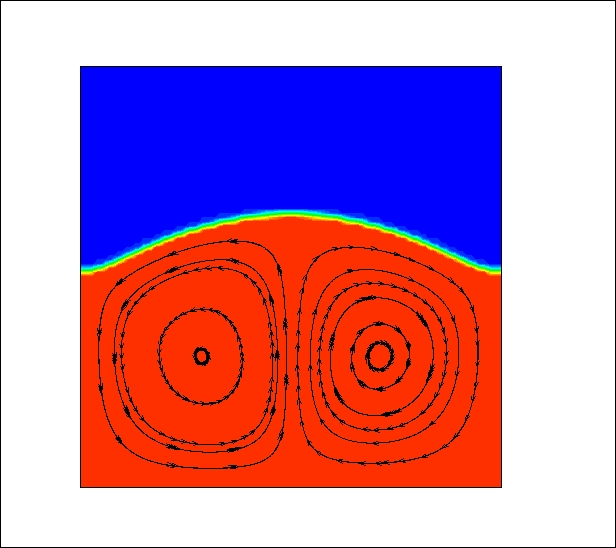} 
\end{subfigure}
\bigskip
\begin{subfigure}{.5\textwidth}
  \centering
   \caption{t = 127 sec}
  \includegraphics[width=.8\linewidth,trim = 2.0cm 2.0cm 2.0cm 2.0cm,clip]{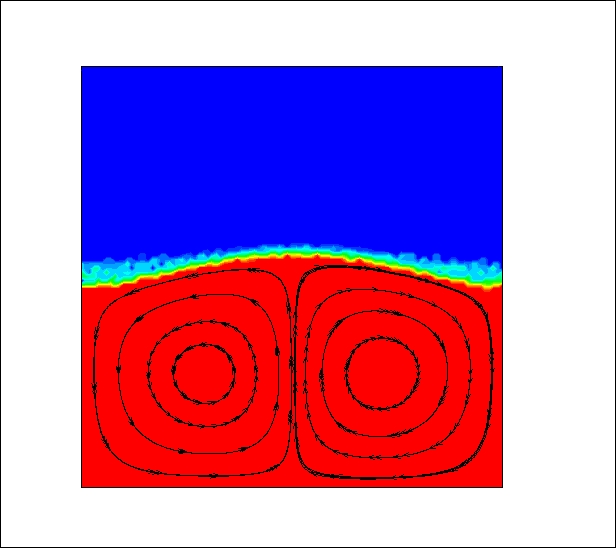}
\end{subfigure}%
\begin{subfigure}{.5\textwidth}
  \centering
  \caption{t = 143 sec}
  \includegraphics[width=.8\linewidth,trim = 2.0cm 2.0cm 2.0cm 2.0cm,clip]{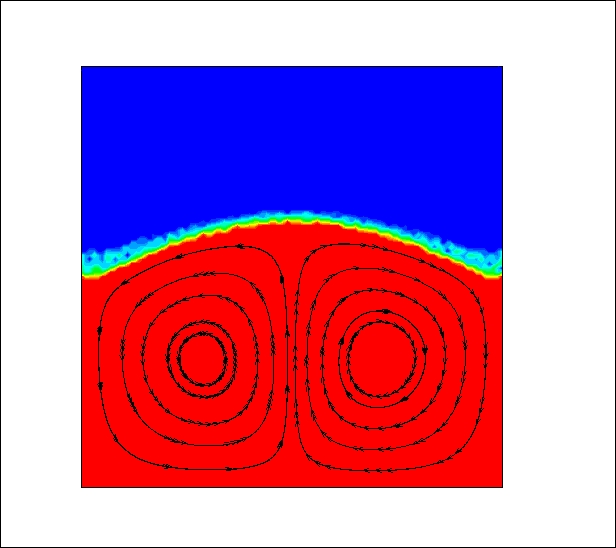}
\end{subfigure}
\bigskip
\begin{subfigure}{.5\textwidth}
  \centering
\includegraphics[width= 1.2\linewidth,trim = 0cm 0cm 0cm 0cm,clip]{Liquid_fraction_bar.jpg}
\end{subfigure}
\caption{Comparison between evolution of the flow field (streamlines) superimposed on the contours of the liquid fractions for pure water ($\phi$=0) (a), (b), and for the case of the NePCM ($\phi_w$ = 10\%), for $T_h$ = 290 K, $T_c$ = 270 K.}
\label{Figliquidbottmcomprassionpurewater}
\end{figure}

We are comparing the flow fields (streamlines) superimposed on the liquid fraction contours for the cases of pure water and NePCM with a mass concentration of particles of 10\%, Ra =$\eval*{$10^{-4}\frac{\rho g \beta(Th - Tc)H^3}{\mu\frac{k}{\rho cp}}$}[\rho = 1.1\times 10^{3}, g = 9.8,\beta = 1.9\times 10^{-4}, Th = 290, Tc = 270, H = 0.005, \mu = 9.18\times 10^{-4}, k = 6.23\times 10^{-1}, cp = 3.8\times 10^{3}][x2] \times 10^4$, as shown in Figure \ref{Figliquidbottmcomprassionpurewater}. The comparison reveals that both cases' flow fields generated by convection are nearly similar\@. However, the flow intensity is higher in the case of pure water compared to NePCM, with a mass concentration of particles of 10\%\@. The solid-liquid interface advances further in the case of pure water compared to NePCM with a mass concentration of particles of 10\% due to the high intensity of the convection for the case of pure water\@. This deceleration of the melting front is also observed in the experiments conducted by Zeng et al. \cite{zeng2013experimental} for the bottom melting for the NePCM\@.\\

\begin{figure}[p!]
\begin{center}
\includegraphics [scale= 0.8, trim = 100 30 0 0,clip]{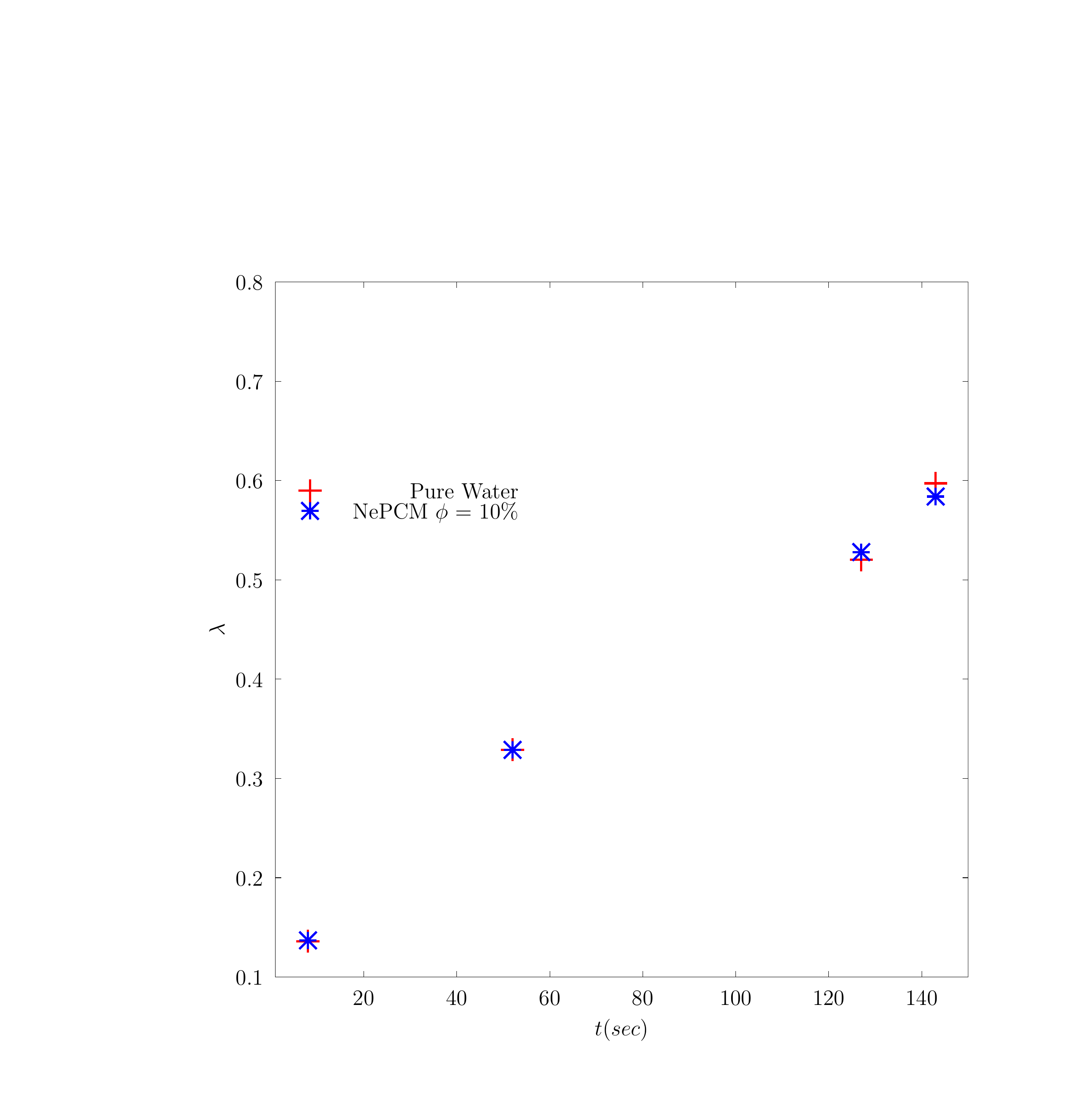}
\end{center}
\caption{Comparison  between the variation of the average liquid fraction for different time intenseness, for the case of pure water, and NePCM $\phi_w$ = 10\%, for  melting from the bottom with $T_h$ = 290 K, $T_c$ = 270 K.}
\label{Figliqave_bootom_melting_2}
\end{figure}

To have a better understanding of the melting performance of the NePCM in the case of bottom melting, we will compare the time evolution of the average liquid fraction with that of water at different time instances, as shown in Figure \ref{Figliqave_bootom_melting_2}\@. In the melting process's early stage, pure water and the NePCM melt at the same rate\@. However, in the late melting stage, pure water melts faster than the NePCM, which can be observed from Figure \ref{Figliquidbottmcomprassionpurewater}, where convection becomes the dominant mechanism controlling the melting process\@. The water melts faster by about \eval*{$(\frac{(0.59727-0.584)}{0.59727})\times 100 $}[x2] \%. \\

\begin{figure}[p!]
\begin{center}
\includegraphics [scale= 0.8, trim = 100 30 0 0,clip]{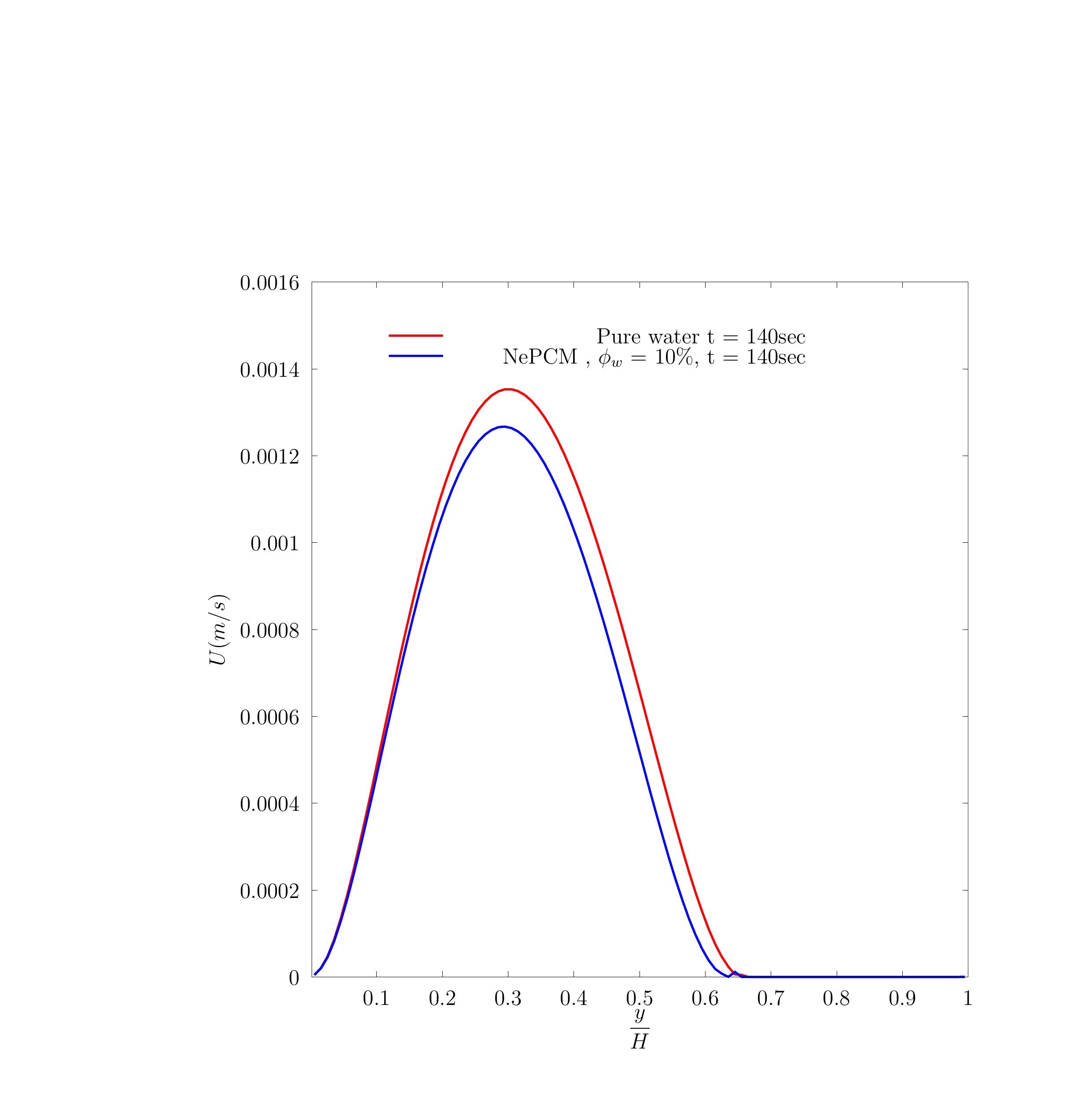}
\end{center}
\caption{The variation of the velocity  along a vertical line that crosses the bottom, and top sides of the cavity at $\dfrac{x}{H}$ = 0.5, for the case of for the case of melting from the bottom for  $T_h$ = 290 K, $T_c$ = 270 K, and $\phi_w$ = 10\%. }
\label{Figvel_line_bootom_melting_2}
\end{figure}

To further clarify our observations, we plot the velocity and temperature along a vertical line extending from the bottom to the top sides of the cavity, located at $\frac{x}{H} = 0.5$@. Figure \ref{Figvel_line_bootom_melting_2} shows the velocity distribution\@. Interestingly, near the bottom wall, the velocity of pure water and the NePCM are equal\@. However, as we move further from the wall, the velocity of the water becomes higher than that of the NePCM due to stronger convection currents\@. After reaching a maximum value, the velocity decreases due to a reduction in convection intensity \@. From the peak point onwards, the velocity of pure water is higher than that of the NePCM\@. \\

\begin{figure}[p!]
\begin{center}
\includegraphics [scale= 0.8, trim = 100 30 0 0,clip]{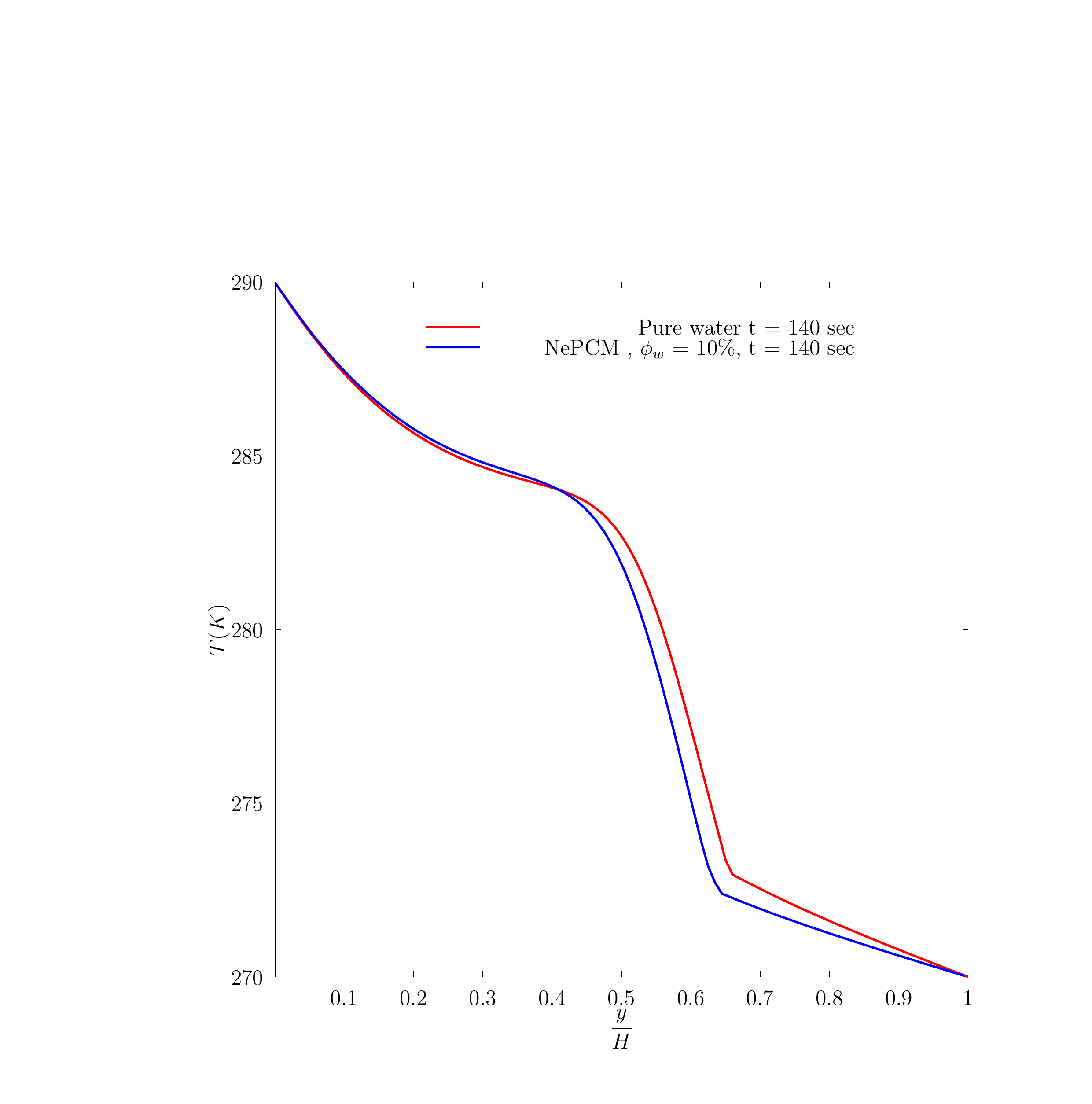}
\end{center}
\caption{The variation of the temperature along a vertical line that crosses the bottom, and top sides of the cavity at $\dfrac{x}{H}$ = 0.5, for the case of for the case of melting from the bottom for  $T_h$ = 290 K, $T_c$ = 270 K, and $\phi_w$ = 10\%. }
\label{Figtemp_line_bootom_melting_2}
\end{figure}

The temperature profiles are depicted in Figure \ref{Figtemp_line_bootom_melting_2}\@. The temperature of the NePCM is slightly higher than that of pure water near the bottom wall, which can be attributed to the higher thermal conductivity of the NePCM\@. However, as we move away from the bottom wall, the temperature of pure water becomes higher than that of the NePCM due to better mixing facilitated by convection\@.     
\subsubsection{effect of the Rayleigh number}

We will increase the Rayleigh number by raising the temperature of the bottom wall to further investigate the natural convection effect on the melting process of the NePCM\@. We will increase the bottom temperature to $T_h$ = 300 K. The corresponding value of the thermal Rayleigh number is  to \eval*{$10^{-4}\frac{\rho g \beta(Th - Tc)H^3}{\mu\frac{k}{\rho cp}}$}[\rho = 1.1\times 10^{3}, g = 9.8,\beta = 2.1\times 10^{-4}, Th = 300,Tc = 270, H = 0.005, \mu =8.9\times 10^{-4}, k =
6.0\times 10^{-1}, cp = 4.179\times 10^{3}][x2] $\times 10^4$ for the case of the pure water, and for the case of NePCM with $\phi_w$ = 10\%, $Ra$ =  $\eval*{$10^{-4}\frac{\rho g \beta(Th - Tc)H^3}{\mu\frac{k}{\rho cp}}$}[\rho = 1.1\times 10^{3}, g = 9.8,\beta = 1.9\times 10^{-4}, Th = 300, Tc = 270, H = 0.005, \mu = 9.18\times 10^{-4}, k = 6.23\times 10^{-1}, cp = 3.8\times 10^{3}][x2] \times 10^4$\@ .  \\

\begin{figure}[p!]
\centering
\begin{subfigure}{.5\textwidth}
 \centering
 \caption{t= 8 sec}
  \includegraphics[width=.8\linewidth, trim =2.0cm 2.0cm 2.0cm 1.0cm,clip ]{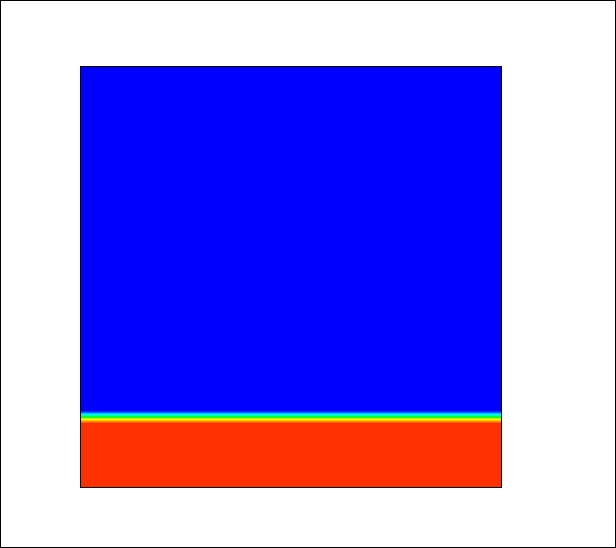}
\end{subfigure}%
\begin{subfigure}{.5\textwidth}
  \centering
  \caption{t = 52 sec}
  \includegraphics[width=.8\linewidth,trim = 2.0cm 2.0cm 2.0cm 2.0cm,clip]{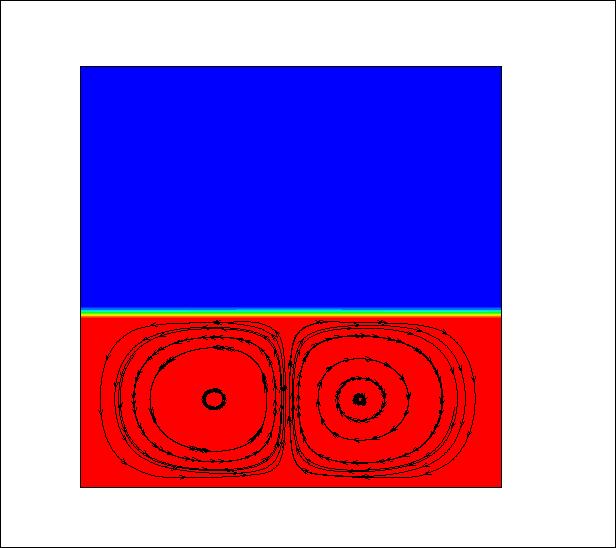}
\end{subfigure}
\bigskip
\begin{subfigure}{.5\textwidth}
  \centering
   \caption{t = 127 sec}
  \includegraphics[width=.8\linewidth,trim = 2.0cm 2.0cm 2.0cm 2.0cm,clip]{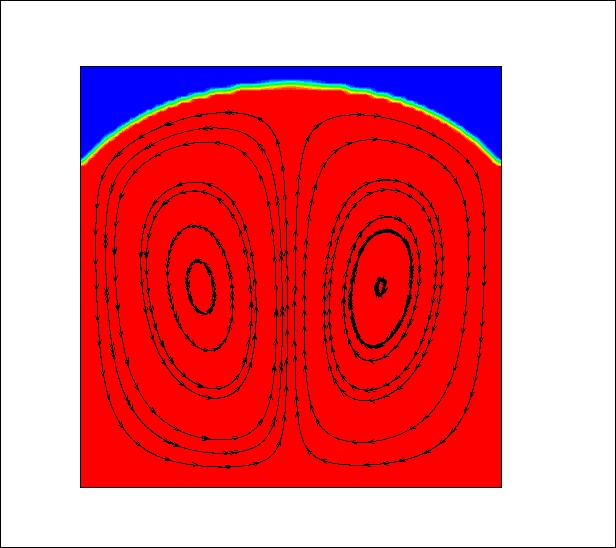}
\end{subfigure}%
\begin{subfigure}{.5\textwidth}
  \centering
  \caption{t = 143 sec}
  \includegraphics[width=.8\linewidth,trim = 2.0cm 2.0cm 2.0cm 2.0cm,clip]{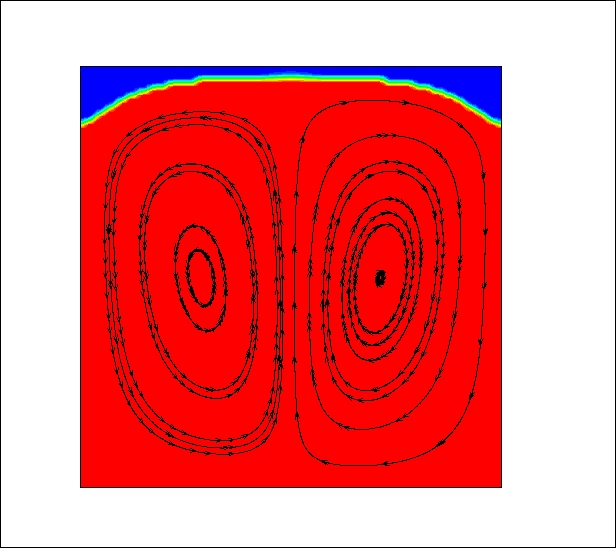}
\end{subfigure}
\bigskip
\begin{subfigure}{.5\textwidth}
  \centering
  \includegraphics[width= 1.2\linewidth,trim = 0cm 0cm 0cm 0cm,clip]{liquid_fraction_top_bar.jpg}
\end{subfigure}
\caption{Development of the flow field (shown by the stream lines) superimposed on the contour of the liquid fraction   for  $T_h$ =300 K, $T_c$ = 270 K, for case of  pure water\@.}
\label{Figliquidbottom3_pure_water}
\end{figure}

The superimposed contours of streamlines and the solid-liquid interface for the case of pure water ($\phi_w$ = 0) are shown in Figure \ref{Figliquidbottom3_pure_water}. At t = 8 sec, the flow is predominantly governed by conduction, with the fluid velocity approximately equal to zero, and the solid-liquid interface being flat. However, at t = 52 sec, the effects of convection start to become apparent, leading to the formation of two symmetric counter-rotating vortices of nearly equal strength.
As time progresses, these two counter-rotating cells occupy a significant portion of the melted water zone, while the solid-liquid interface takes on a curved and symmetric shape, in contrast to the initial flat shape during the early stages of the melting process. This change in shape of the solid-liquid interface is a result of the substantial increase in convection strength within the cells.
Both cases of water, with melting occurring from the bottom with $Ra$ values of $3.74 \times 10^4$ and $7.31 \times 10^4$ respectively, have nearly  identical, flow fields and solid-liquid interface evolution\@. The only difference is that the latter case experiences a faster melt rate than the former\@.\\

\begin{figure}[p!]
\centering
\begin{subfigure}{.5\textwidth}
 \centering
 \caption{t= 8 sec}
  \includegraphics[width=.8\linewidth, trim =2.0cm 2.0cm 2.0cm 1.0cm,clip ]{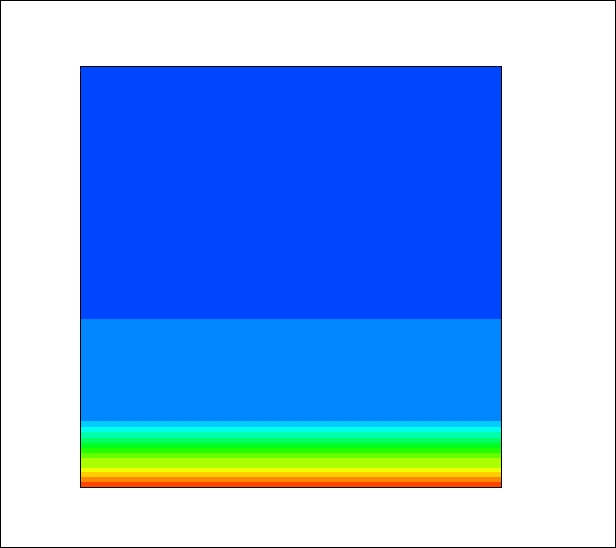}
\end{subfigure}%
\begin{subfigure}{.5\textwidth}
  \centering
  \caption{t = 52 sec}
  \includegraphics[width=.8\linewidth,trim = 2.0cm 2.0cm 2.0cm 2.0cm,clip]{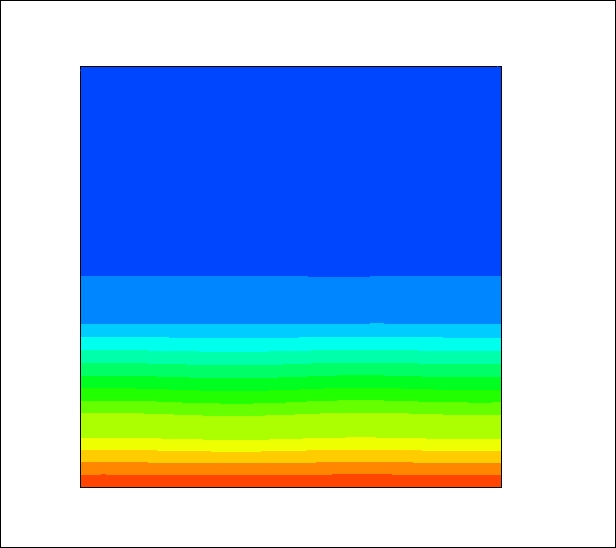}
\end{subfigure}
\bigskip
\begin{subfigure}{.5\textwidth}
  \centering
   \caption{t = 127 sec}
  \includegraphics[width=.8\linewidth,trim = 2.0cm 2.0cm 2.0cm 2.0cm,clip]{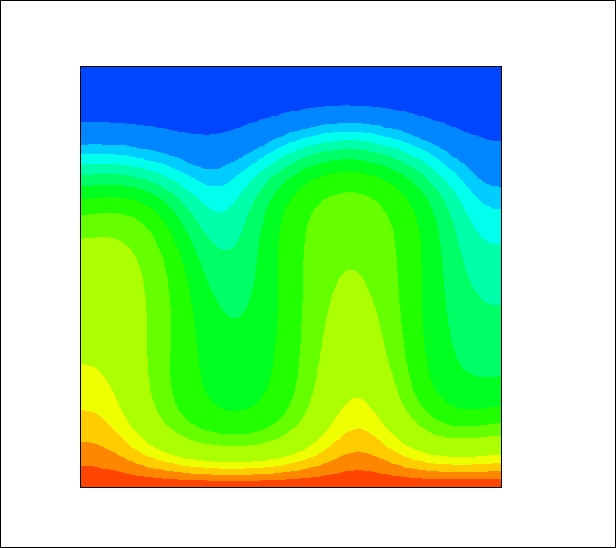}
\end{subfigure}%
\begin{subfigure}{.5\textwidth}
  \centering
  \caption{t = 143 sec}
  \includegraphics[width=.8\linewidth,trim = 2.0cm 2.0cm 2.0cm 2.0cm,clip]{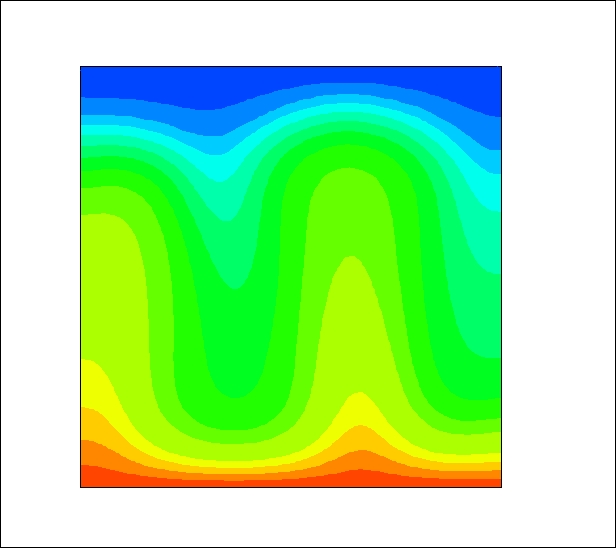}
\end{subfigure}
\bigskip
\begin{subfigure}{.5\textwidth}
  \centering
  \includegraphics[width= 1.2\linewidth,trim = 0cm 0cm 0cm 0cm,clip]{Temperature_legend1.jpg}
\end{subfigure}
\caption{The evolution of temperature  profiles  for the case of melting from the bottom for  $T_h$ = 300 K, $T_c$ = 270 K, and $\phi_w$ = 10\%.}
\label{Figtempebottom_melting_3}
\end{figure}

For the case of the NePCM with $\phi_w$ = 10\% and $Ra$ = $5.61 \times 10^4$, the superimposed contours of streamlines and the solid-liquid interface are depicted in Figure \ref{Figtempebottom_melting_3}. At t = 8 sec, the solid-liquid interface is flat, and the melting process is primarily governed by conduction, similar to the pure water case.
However, as time progresses, the flow field evolves to consist of three vortexes. The middle vortex rotates counter-clockwise, while the other two surrounding vortexes rotate clockwise. The left vortex exhibits a higher strength compared to the remaining two.
As time further elapses to t = 127 sec and 143 sec, the symmetry of the solid-liquid interface is disrupted, and the three vortexes grow in size. The melting near the right vertical wall occurs at a faster rate, whereas the growth of the solid-liquid interface near the left vertical wall is considerably slower.
The breaking of symmetry in the solid-liquid interface can be attributed to the generation of thermosolutal convection resulting from the redistribution of the particles during the melting process. This physical mechanism plays a crucial role in the observed phenomena, and thus using single phase model  to model the melting process of the NePCM is an oversimplification that leads to significant error in the predictions\@.   Also, the difference in the flow fields between the pure water and the NePCM is responsible for the clear deceleration of the melting process of the NePCM compared to the water, not the increase of the viscosity as several investigations show in the literature {show references}\@.  \\

\begin{figure}[p!]
\centering
\begin{subfigure}{.5\textwidth}
 \centering
 \caption{t= 8 sec}
  \includegraphics[width=.8\linewidth, trim =2.0cm 2.0cm 2.0cm 1.0cm,clip ]{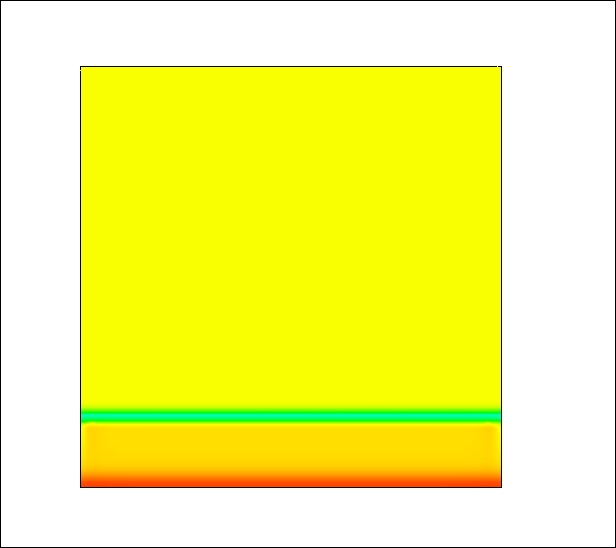}
\end{subfigure}%
\begin{subfigure}{.5\textwidth}
  \centering
  \caption{t = 52 sec}
  \includegraphics[width=.8\linewidth,trim = 2.0cm 2.0cm 2.0cm 2.0cm,clip]{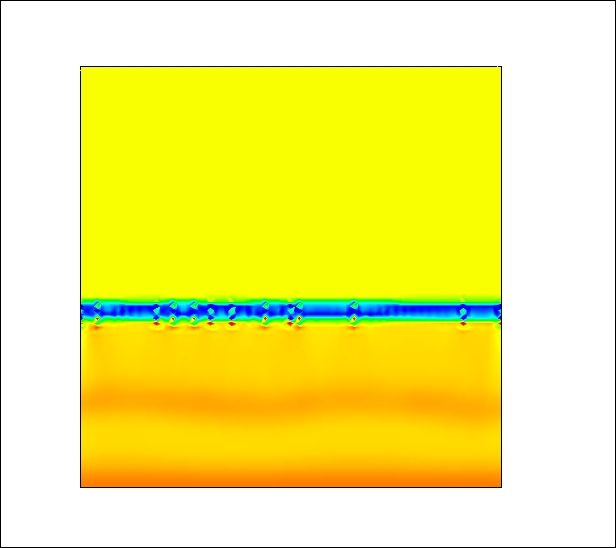}
\end{subfigure}
\bigskip
\begin{subfigure}{.5\textwidth}
  \centering
   \caption{t = 127 sec}
  \includegraphics[width=.8\linewidth,trim = 2.0cm 2.0cm 2.0cm 2.0cm,clip]{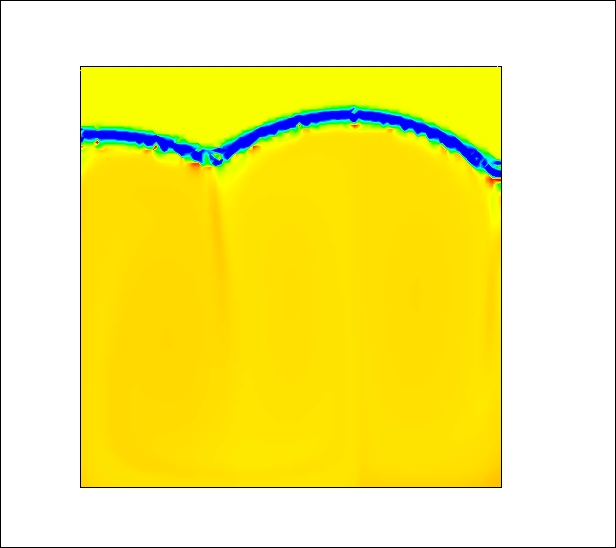}
\end{subfigure}%
\begin{subfigure}{.5\textwidth}
  \centering
  \caption{t = 143 sec}
  \includegraphics[width=.8\linewidth,trim = 2.0cm 2.0cm 2.0cm 2.0cm,clip]{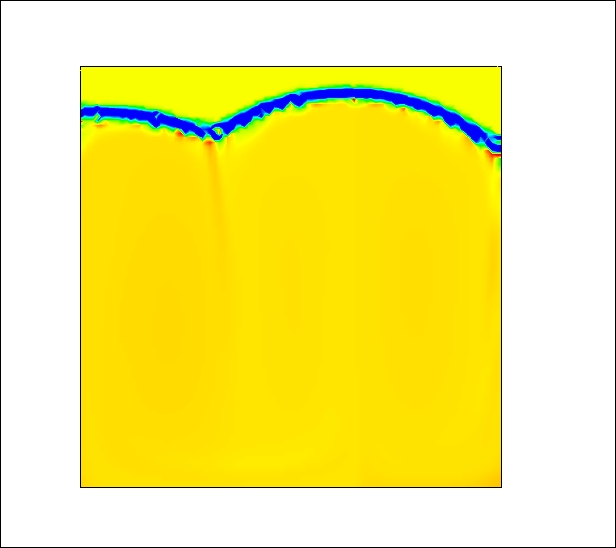}
\end{subfigure}
\bigskip
\begin{subfigure}{.5\textwidth}
  \centering
  \includegraphics[width= 1.2\linewidth,trim = 0cm 0cm 0cm 0cm,clip]{Concentration_bar.jpg}
\end{subfigure}
\caption{The evolution of concentration profiles of nano-particles for the case of melting from the bottom for  $T_h$ = 300 K, $T_c$ = 270 K, and $\phi_w$ = 10\%.}
\label{Figconbottommelting3}
\end{figure}

To further investigate the non-symmetric shape of the solid-liquid interface we observed in Figure \ref{Figliquidbottom3_pure_water}, we plotted the particle redistribution  as shown in Figure \ref{Figconbottommelting3} \@. For $t = 8$ sec and $t = 52$ sec, we observe that the particles in the melting zone are distributed in alternating high and low-concentration layers of particles \@. When the thermosolutal convection starts, the particles are redistributed along the melted portion of the NePCM. Even though the particles are about  \eval*{$\frac{\rho_p}{\rho_f}$} [\rho_p = 8954,\rho_f = 1000] [x2] times heavier than the water\@. The particles are not distributed evenly, and a large portion of them accumulates at the left wall of the cavity, the accumulation of particles at the left wall explains the slowing down of the solid-liquid interface in this particular location of the melting zone \@. Overall, the non-uniform distribution of the particles leads to the non-symmetric solid-liquid interface observed \@.\\

\begin{figure}[p!]
\centering
\begin{subfigure}{.5\textwidth}
 \centering
 \caption{t= 8 sec}
  \includegraphics[width=.8\linewidth, trim =2.0cm 2.0cm 2.0cm 1.0cm,clip ]{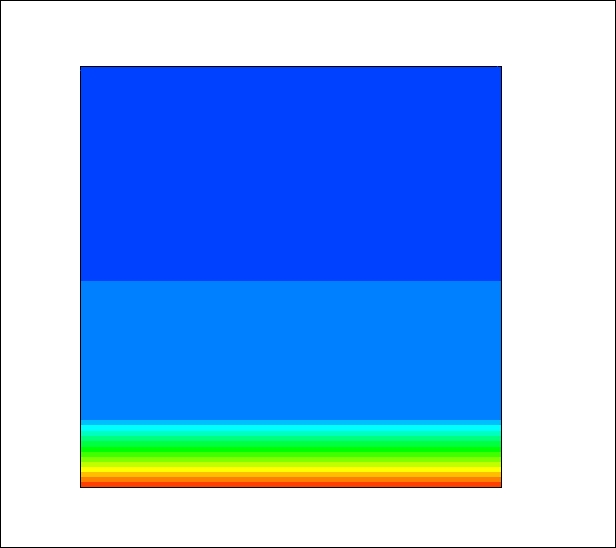}
\end{subfigure}%
\begin{subfigure}{.5\textwidth}
  \centering
  \caption{t = 52 sec}
  \includegraphics[width=.8\linewidth,trim = 2.0cm 2.0cm 2.0cm 2.0cm,clip]{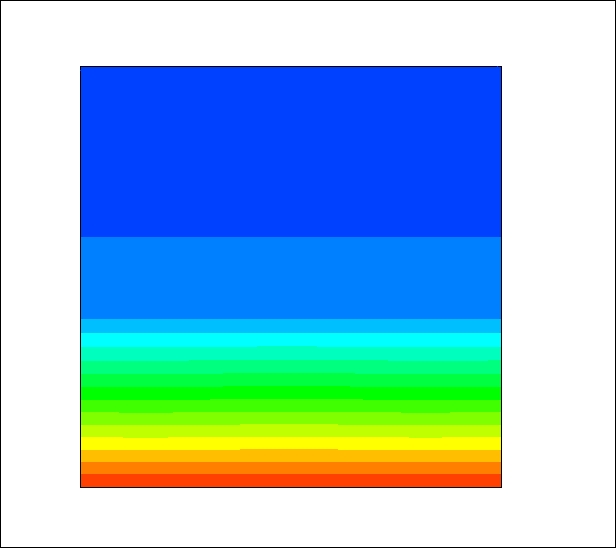}
\end{subfigure}
\bigskip
\begin{subfigure}{.5\textwidth}
  \centering
   \caption{t = 127 sec}
  \includegraphics[width=.8\linewidth,trim = 2.0cm 2.0cm 2.0cm 2.0cm,clip]{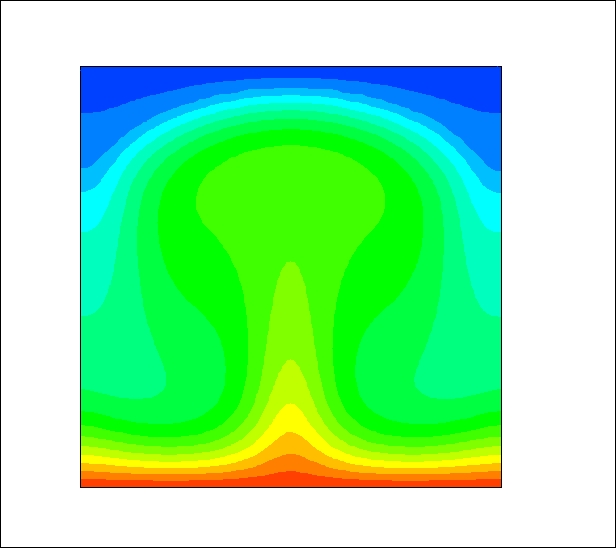}
\end{subfigure}%
\begin{subfigure}{.5\textwidth}
  \centering
  \caption{t = 143 sec}
  \includegraphics[width=.8\linewidth,trim = 2.0cm 2.0cm 2.0cm 2.0cm,clip]{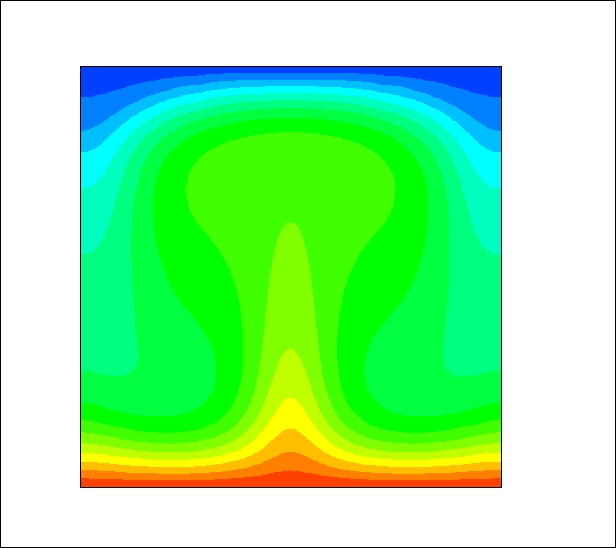}
\end{subfigure}
\bigskip
\begin{subfigure}{.5\textwidth}
  \centering
  \includegraphics[width= 1.2\linewidth,trim = 0cm 0cm 0cm 0cm,clip]{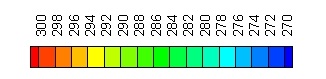}
\end{subfigure}
\caption{The evolution of temperature  profiles  for the case of melting from the bottom for  $T_h$ = 300K, $T_c$ = 270 K, for the case of pure water.}
\label{Figtempebottom3_pure_water}
\end{figure}

The temperature distribution for the case of pure water with $Ra = 7.3 \times 10^4$ is depicted in Figure \ref{Figtempebottom3_pure_water}\@. At the beginning of the melting process, the temperature profile indicates conduction as the dominant heat transfer mechanism\@. However, thermal convection becomes the dominant factor influencing the temperature profile as time progresses\@. The warmest temperature zone is situated along the vertical line crossing the bottom wall of the system\@.
On the other hand, for the NePCM case with $Ra = 5.61 \times 10^4$, the temperature profiles are illustrated in Figure \ref{Figtempebottom_melting_3}\@. Initially, during the early stages of melting, the temperature profiles resemble those observed for pure water\@. However, as thermosolutal convection becomes more prominent, the temperature profiles become asymmetric\@. In the case of the NePCM, two warm zones are observed, one near the left bottom corner of the duct and the other to the right of the duct's center\@. These warm zones are closely related to the locations where the particles accumulate the most\@. The flow, temperature, and concentration fields are strongly coupled\@. Neglecting the transport of particles would lead to significant overestimations that do not reflect reality\@.\\

\begin{figure}[p!]
\begin{center}
\includegraphics [scale= 0.8, trim = 100 30 0 0,clip]{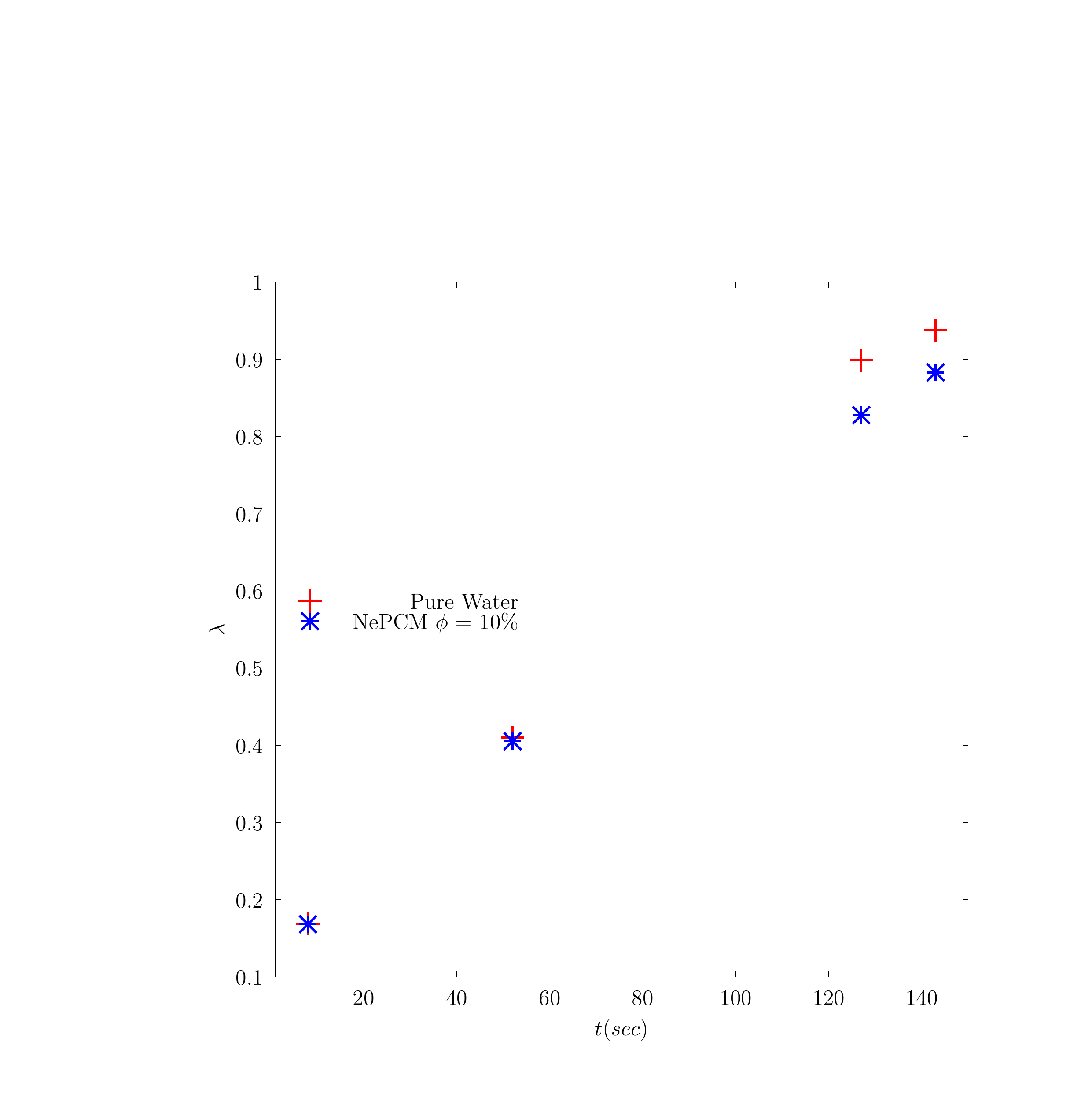}
\end{center}
\caption{Comparison  between the variation of the average liquid fraction for different time intenseness, for the case of pure water, and NePCM $\phi_w$ = 10\%, for  melting from the top with $T_h$ = 300 K, $T_c$ = 270 K. }
\label{Figtimeaveliq_bottommelting3}
\end{figure}

We will compare the liquid fraction time evolution  for the case of pure  water with that of the NePCM with mass fraction of particles ($\phi_w$ ) = 10\% for case of bottom wall temperature ($T_h$) = 300 K, and the top wall temperature ($T_c$) = 270 K, as shown in Figure  \ref{Figtimeaveliq_bottommelting3}\@. At the early stage of melting the pure water and the NePCM are melting with the same rate, however when the convection starts, the water melts faster than the NePCM\@. The water melts faster by about \eval*{$(\frac{(0.9376-0.883)}{0.9376})\times 100 $}[x2] \% \@.  If we check the thermal Rayleigh numbers  for the case of pure water, and the NePCM we will find that they differ by \eval*{$\frac{(7.3-5.61)}{7.3}\times 100$} [1]  \%, the difference in the Rayleigh number is much higher than the  difference in the melting rate slowdown, which suggest that the water most melted by higher rate than 5.82\% which leads us to conclude that the thermal Rayleigh number, is not the only parameter that is contributing to the deceleration of the solid-liquid interface, but the combined thermosolutal convection is the leading parameter that controls the melting process of the NePCM, as will show with extra evidence later \@. If the thermal convection is not the leading parameter of controlling the melting process of the NePCM, thus  increasing the viscosity of the NePCM due to the suspension of the particles, is not the only reason for the deceleration of melting as it was thought before \cite{fan2014heat,dhaidan2015melting}\@. Another observation is that if we increase the temperature of  bottom  wall decseleration of the melting rate of NePCM will increase compared to water\@. \\

We will further analyse the cases of the pure water, and NePCM  for the case of melting from the bottom side for the thermal boundary conditions of $T_h$ = 300 K, and $T_c$ = 270 K, by plotting different quantities along a vertical line that it is extended from the middle of the bottom wall to the middle of the top wall\@. \\

\begin{figure}[p!]
\begin{center}
\includegraphics [scale= 0.8, trim = 100 30 0 0,clip]{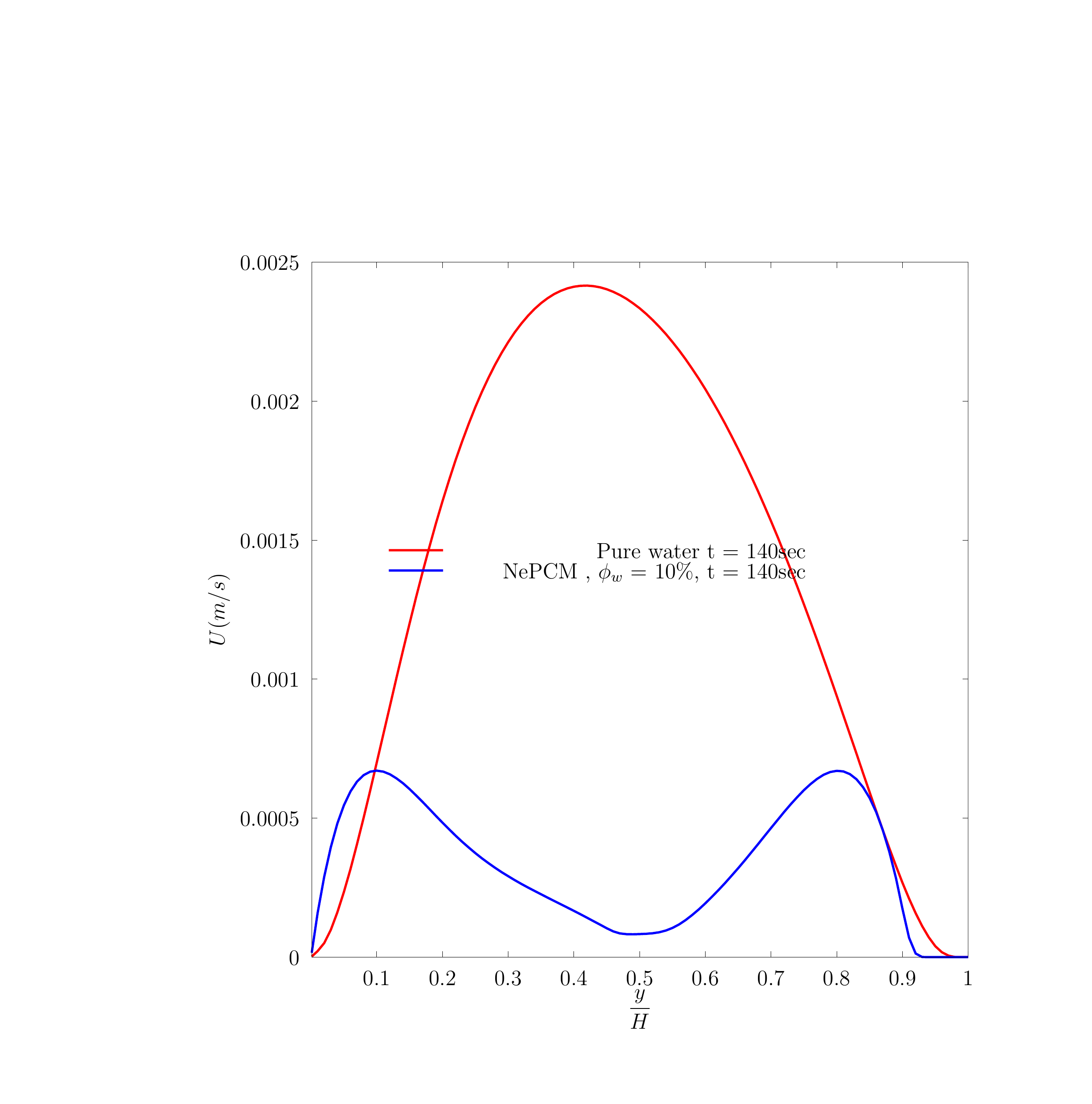}
\end{center}
\caption{ Comparison between the  variation of the velocity  along a vertical line that crosses the bottom, and top sides of the cavity at $\dfrac{x}{H}$ = 0.5, for the case of for the case of melting from the bottom for  $T_h$ = 300 K, $T_c$ = 270 K, for pure water, and NePCM with $\phi_w$ = 10\%. }
\label{Figvel_line_bootom_melting_3}
\end{figure}

\begin{figure}[p!]
\begin{center}
\includegraphics [scale= 0.8, trim = 100 30 0 0,clip]{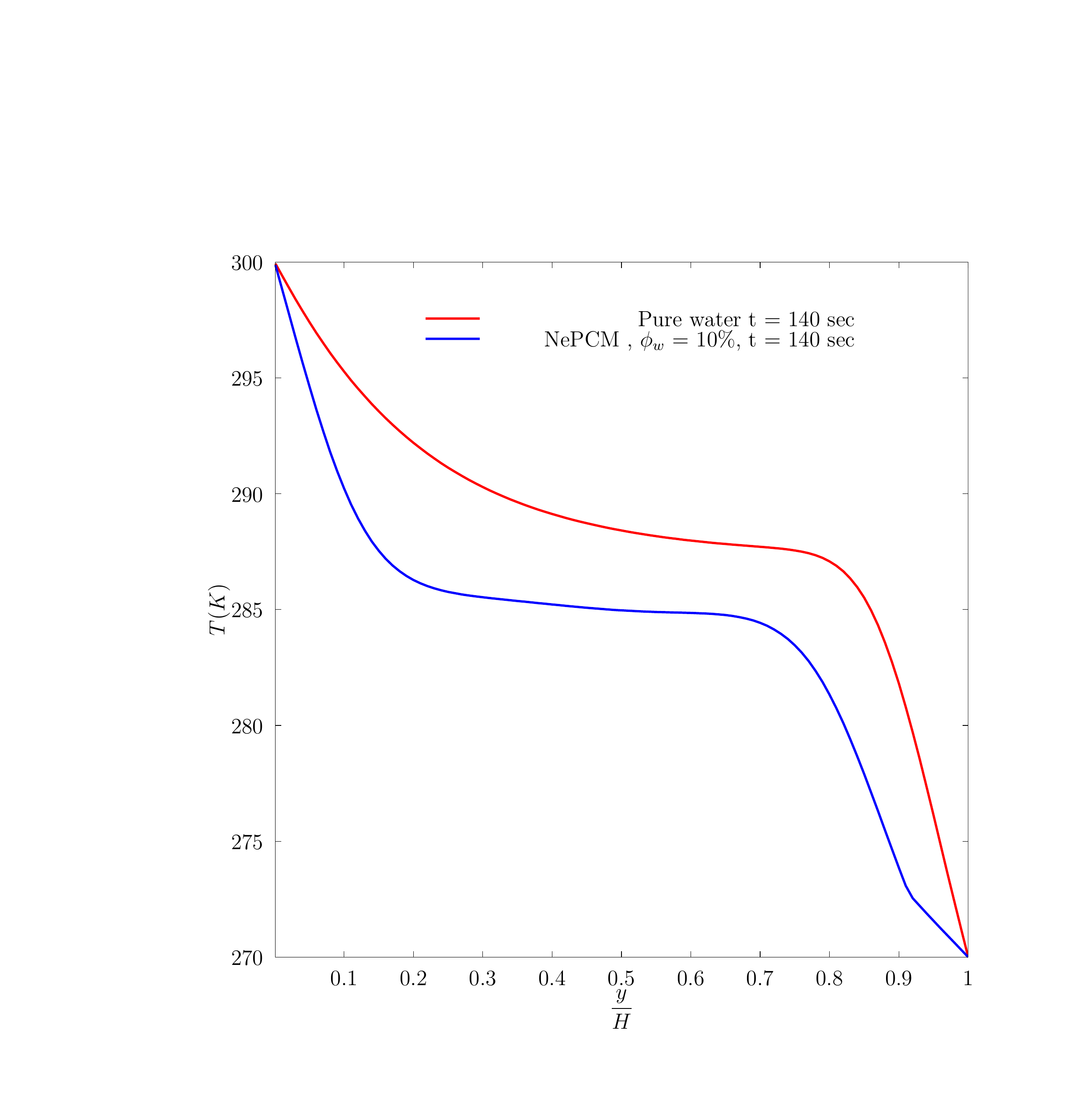}
\end{center}
\caption{Comparison between the variation of the  temperature along a vertical line that crosses the bottom, and top sides of the cavity at $\dfrac{x}{H}$ = 0.5, for the case of for the case of melting from the bottom for  $T_h$ = 300 K, $T_c$ = 270 K, for pure water and NePCM with $\phi_w$ = 10\%. }
\label{Figtemp_line_bootom_melting_3}
\end{figure}

\begin{figure}[p!]
\begin{center}
\includegraphics [scale= 0.8, trim = 100 30 0 0,clip]{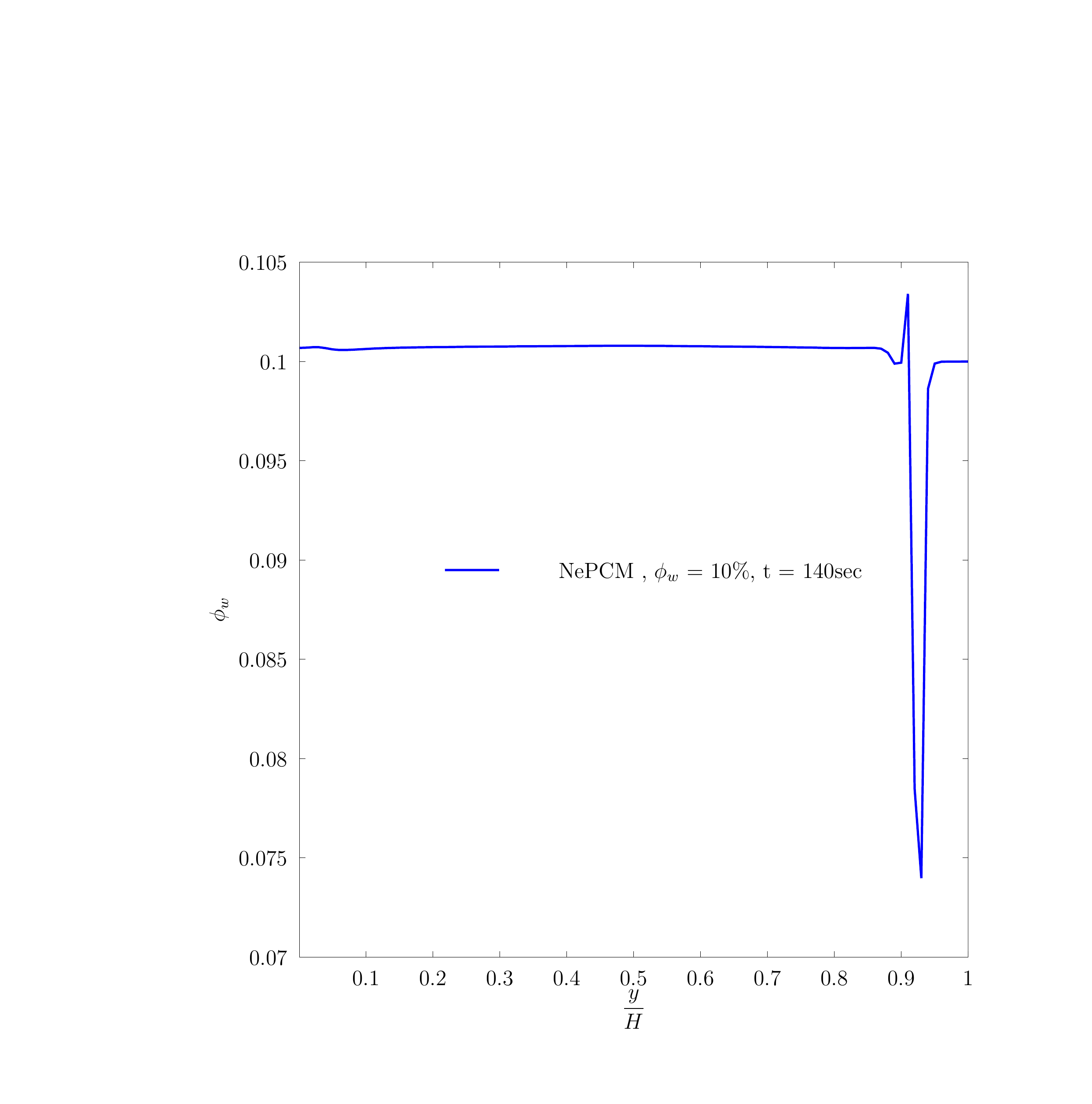}
\end{center}
\caption{The variation of the mass fraction of the particles  along a vertical line that crosses the bottom, and top sides of the cavity at $\dfrac{x}{H}$ = 0.5, for the case of for the case of melting from the bottom for  $T_h$ = 300 K, $T_c$ = 270 K, for  NePCM with $\phi_w$ = 10\%. }
\label{Figcon_line_bootom_melting_3}
\end{figure}

Figure \ref{Figvel_line_bootom_melting_3} the velocity profiles for the pure water, and the NePCM, for the case of the pure water the velocity profile is nearly symmetric, with a local maximum\@. While, for the case of the NePCM  the velocity profile is distorted due to the  thermosolutal convection effect, and the velocity profile consists of two local maxima points, and one local minimum around the middle the hight of the duct\@. Beyond the distortion of the velocity profile the thermosolutal convection effects also the magnitude of the velocity significantly, for example the maximum velocity for the case of pure water is about 5 times greater than the maximum velocity of the NePCM \@. The concentration profile for the case of the NePCM is shown in Figure \ref{Figcon_line_bootom_melting_3} \@. There is a high concentration of particles near the bottom wall  for distance of about $\dfrac{y}{H}$ $\simeq$ 0.1, then the particles have a uniform concentration throughout the melting  zone, with concentration higher than the initial concentration of 10\% \@. Near the solid-liquid interface, we observe a local maximum value, then a sharp decrease in the concentration of the particles in the mushy zone. In the non-melted part of the NePCM, the concentration of the particles retains its initial value\@. Figure \ref{Figtemp_line_bootom_melting_3} shows the variation of the temperature. For the case of water, in which the melting process is controlled purely by thermal convection, the mixing process is better compared to NePCM, which is reflected in the temperature profile, in which the water temperature is higher than that of the NePCM along the vertical line\@. Even in the region near the wall where there is an accumulation of nanoparticles, which creates a zone of higher thermal conductivity, we do not observe an increase in the values of the temperature of the NePCM compared to that of the water\@. \\

\begin{figure}[p!]
\begin{center}

\includegraphics [scale= 0.8, trim = 100 30 0 0,clip]{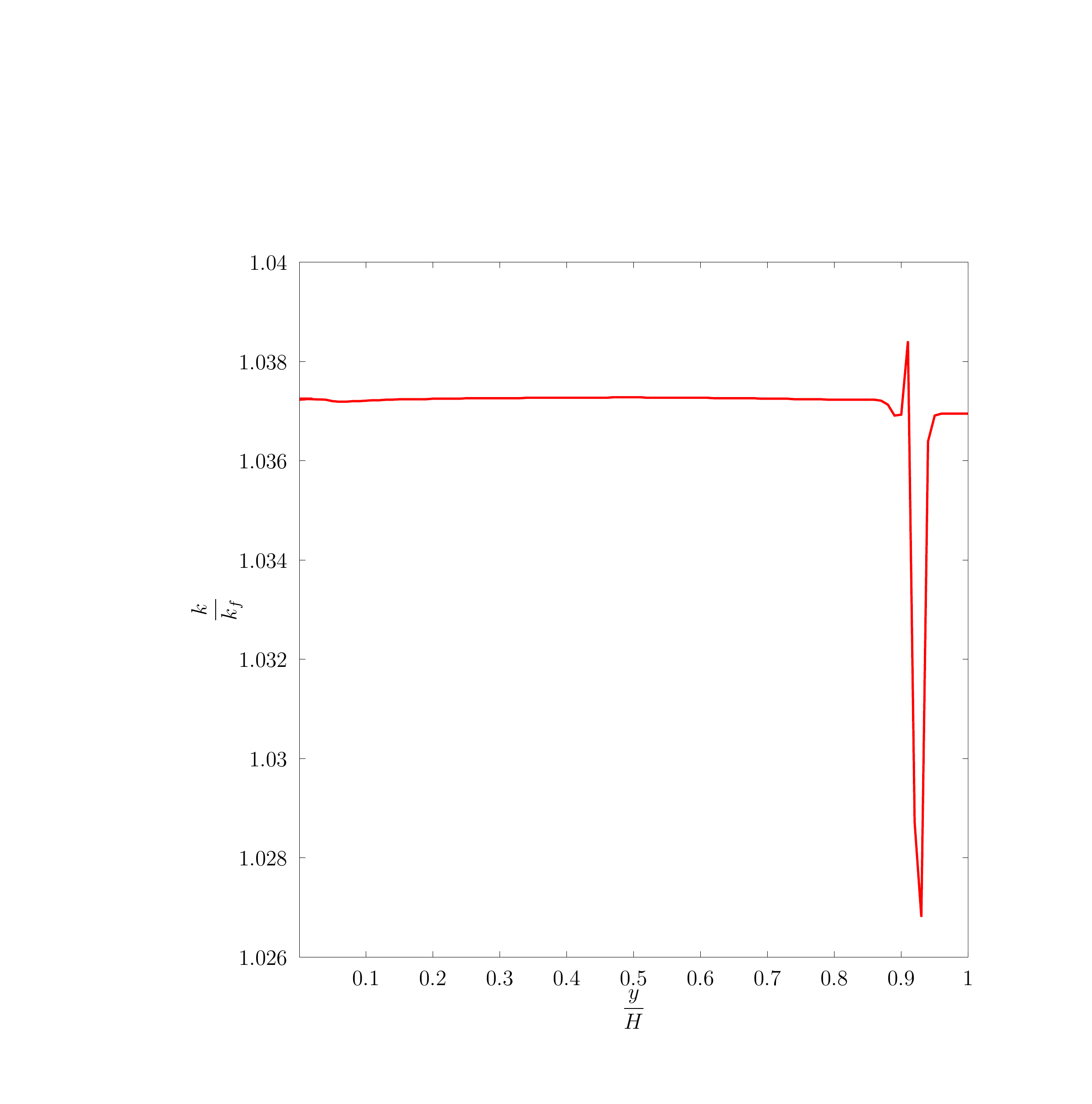}
\end{center}
\caption{The variation of the ratio between the thermal conductivity of the NePCM and that of pure water  along a vertical line that crosses the bottom, and top sides of the cavity at $\dfrac{x}{H}$ = 0.5, for the case of for the case of melting from the bottom for  $T_h$ = 300 K, $T_c$ = 270 K, for  NePCM with $\phi_w$ = 10\%. }
\label{Figthermal_conducvity_line_bootom_melting_3}
\end{figure}

\begin{figure}[p!]
\begin{center}
\includegraphics [scale= 0.8, trim = 100 30 0 0,clip]{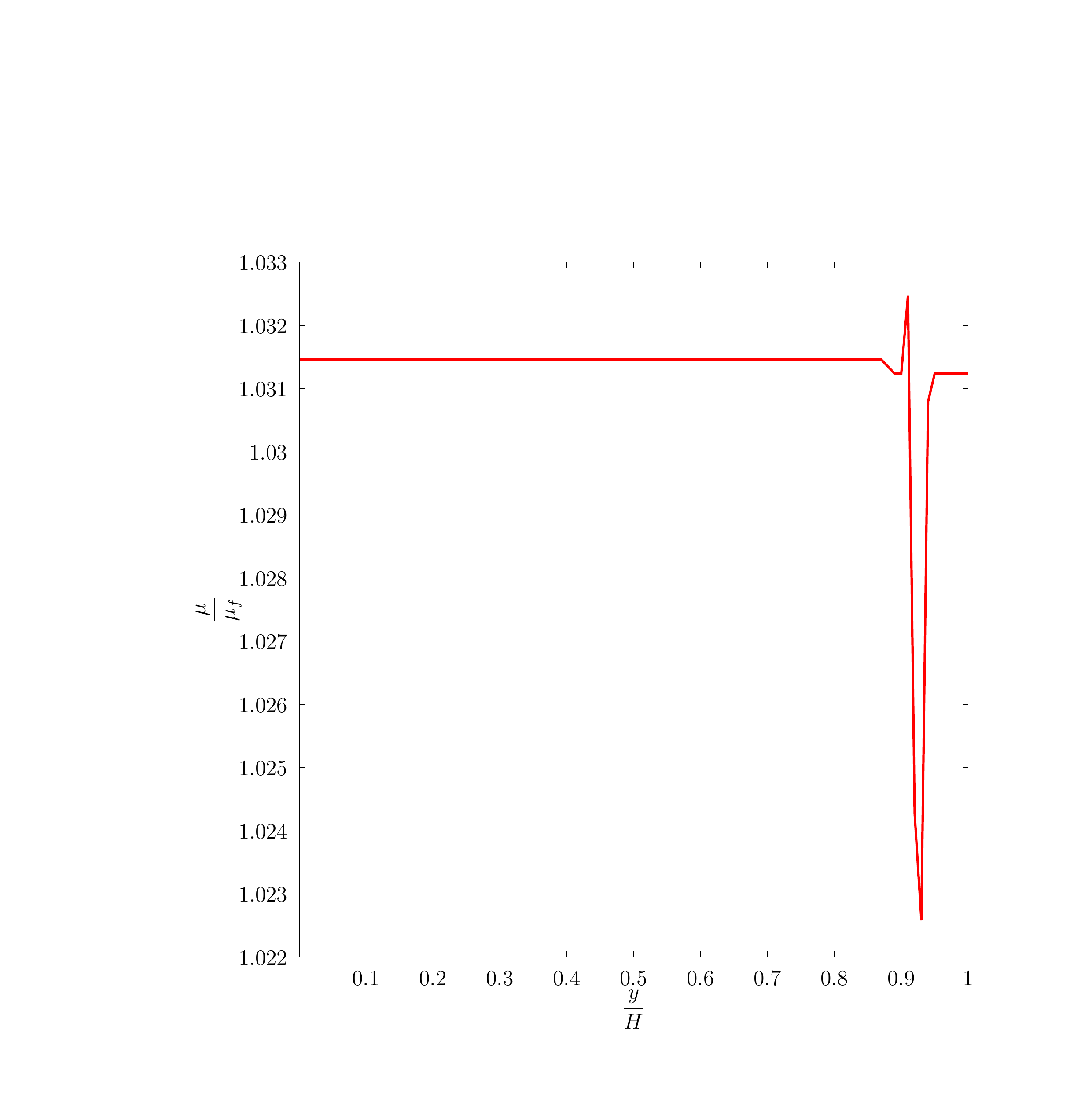}
\end{center}

\caption{The variation of the ratio between the viscosity  of the NePCM and that of pure water  along a vertical line that crosses the bottom, and top sides of the cavity at $\dfrac{x}{H}$ = 0.5, for the case of for the case of melting from the bottom for  $T_h$ = 300 K, $T_c$ = 270 K, for  NePCM with $\phi_w$ = 10\%. }
\label{Figviscousity_line_bootom_melting_3}
\end{figure}

\begin{figure}[p!]
\begin{center}
\includegraphics [scale= 0.8, trim = 100 30 0 0,clip]{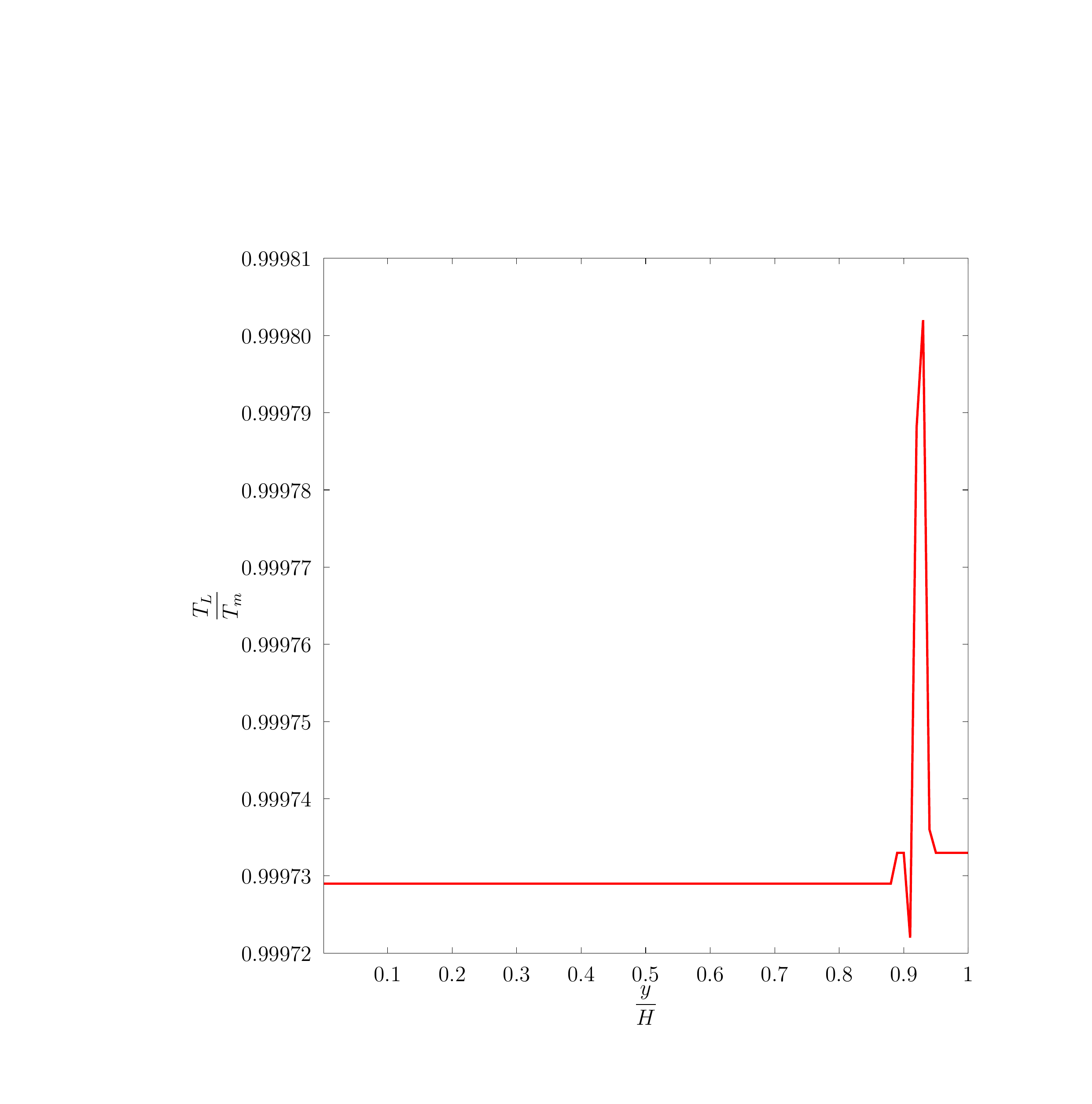}
\end{center}
\caption{The variation of the ratio between the liquid phase melting temperature    of the NePCM and that of pure water  along a vertical line that crosses the bottom, and top sides of the cavity at $\dfrac{x}{H}$ = 0.5, for the case of for the case of melting from the bottom for  $T_h$ = 300 K, $T_c$ = 270 K, for  NePCM with $\phi_w$ = 10\%. }
\label{Figmelting_temperature_line_bootom_melting_3}
\end{figure}

Figure \ref{Figthermal_conducvity_line_bootom_melting_3} shows the variation of the thermal conductivity   ratio of the NePCM to that of the pure water, in the melted region of the NePCM the thermal conductivity is about 3.7\% higher than that of the water, this type of enhancement is consistent  with the findings of global thermal conductivity benchmark exercises for nanofluids \cite{buongiorno2009benchmark}\@. In Figure \ref{Figviscousity_line_bootom_melting_3} we  show the ratio of the viscosity of the NePCM to that of water, the viscosity increase is about 3.3\% which is less than the increase of the thermal conductivity, the increase in the viscosity is consistent with the  international banshmark exercises of the viscosity of stable nanofluids \cite{venerus2010viscosity}\@. The viscosity of the NePCM did not significantly increased due to the redistribution of the particles, and thus is not the main reason for the significant deceleration of the melting rate of the NePCM compared to pure water that is observed\@. We plot also the ratio between the melting temperature of the NePCM, and that of pure water, as shown in Figure \ref{Figmelting_temperature_line_bootom_melting_3}, the two melting temperatures are nearly equal, which shows that the redistribution of the particles did not play a significant role on alerting the Liquids temperature, and thus the liquids temperature variation did not play a significant role in the deceleration of the solid-liquid interface observed for the case of the NePCM compared to that of water\@. \\
\subsection{effect of the segregation coefficient  }
\begin{figure}[p!]
\begin{center}
\includegraphics [scale= 0.8, trim = 100 30 0 0,clip]{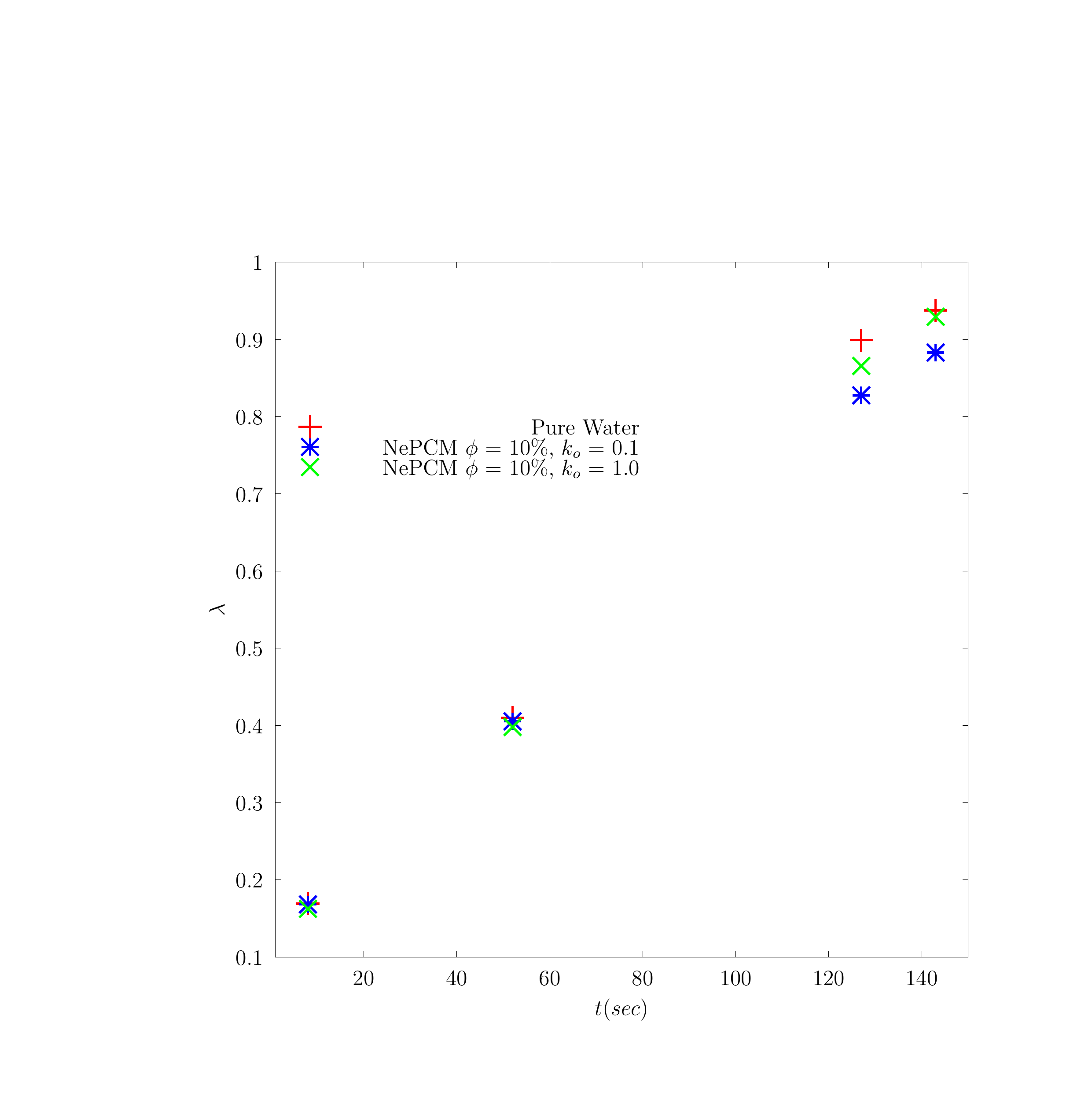}
\end{center}
\caption{Comparison  between the variation of the average liquid fraction for different time intenseness, for the case of pure water, and NePCM $\phi_w$ = 10\%, for  melting from the top with $T_h$ = 300 K, $T_c$ = 270 K with different segregation coefficients.}
\label{Figtimeaveliq_bottommelting3_k}
\end{figure}

\begin{figure}[p!]
\centering
\begin{subfigure}{.5\textwidth}
 \centering
 \caption{Pure Water}
  \includegraphics[width=.8\linewidth, trim =2.0cm 2.0cm 2.0cm 1.0cm,clip ]{Liquid_fraction_143_pure_water_bottom_melting_3.jpg}
\end{subfigure}%
\hfill
\begin{subfigure}{.5\textwidth}
  \centering
   \caption{NePCM $k_o$ = 1.0}
  \includegraphics[width=.8\linewidth,trim = 2.0cm 2.0cm 2.0cm 2.0cm,clip]{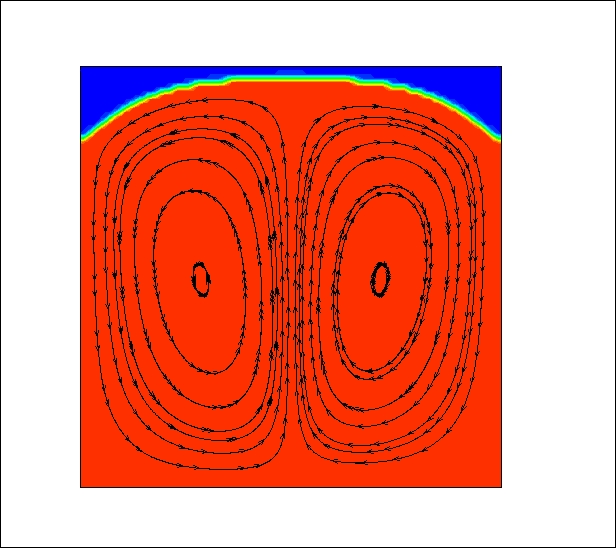}
\end{subfigure}%
\hfill
\begin{subfigure}{.5\textwidth}
  \centering
  \caption{NePCM $k_o$ = 0.1}
  \includegraphics[width=.8\linewidth,trim = 2.0cm 2.0cm 2.0cm 2.0cm,clip]{4.jpg}
\end{subfigure}
\bigskip

\begin{subfigure}{1.0\textwidth}
  \centering
  \includegraphics[width= .7\linewidth,trim = 0cm 0cm 0cm 0cm,clip]{liquid_fraction_top_bar.jpg}
\end{subfigure}
\caption{Development of the flow field (shown by the stream lines) superimposed on the contour of the liquid fraction   for  $T_h$ =300 K, $T_c$ = 270 K, for case of pure water, and  NePCM with $\phi_w$ = 10\%, $t $ = 142 sec, and different segregation coefficients \@.}
\label{Figliquidbottom3different_k}
\end{figure}

To gain more substantial confidence in our conclusion regarding the dominant role of thermosolutal convection in retarding the development of the NePCM melting process compared to pure water, we will set the segregation coefficient to a value of one ($k_0$ = 1.0). This choice signifies that particle concentration will remain uniform within the cavity, indicating that particles will not interact with the solid-liquid interface, or with fluid it self\@. This scenario, represented by $k_0$ = 1.0, aligns with the single-phase models commonly referenced in existing literature\@. As demonstrated in Figure \ref{Figtimeaveliq_bottommelting3_k}, the case where particle redistribution is absent ($k_0$ = 1.0) shows that the NePCM melts at a rate equivalent to that of pure water, consistent with the predictions of many models found in the literature \cite{feng2015numerical}\@.
However, when particle redistribution is present, as in the case of the NePCM with $k_0$ = 0.1, the melting process of the NePCM is notably slower than that of pure water, which is in line with the majority of experimental findings \cite{fan2014heat}\@. The observation depicted in Figure \ref{Figtimeaveliq_bottommelting3_k} strongly supports the notion that particle redistribution by thermosolutal convection serves as the primary factor responsible for the deceleration of the NePCM melting process rather than an increase in viscosity\@. Particle redistribution significantly influences both the flow structure and the shape of the solid-liquid interface, as illustrated in Figure \ref{Figliquidbottom3different_k}\@. In the NePCM scenario with $k_o$=1.0, the flow structure closely resembles that of pure water, featuring two counter-rotating flow cells, distinct from the three-cell flow structure observed with $k_o$ =0.1. This highlights the essential role of particle redistribution in NePCM's melting process and its overall performance\@. Also, both the NePCM with segregation coefficient of 1.0, and 0.1, have the same thermal  Rayleigh number = $5.61\times 10^4$ however, their flow fields, and melting behaviour is different\@. Which shows that thermal Rayleigh number is not the controlling non-dimensional number for the melting process of the NePCM\@.     \\

\begin{figure}[p!]
\begin{center}
\includegraphics [scale= 0.8, trim = 100 30 0 0,clip]{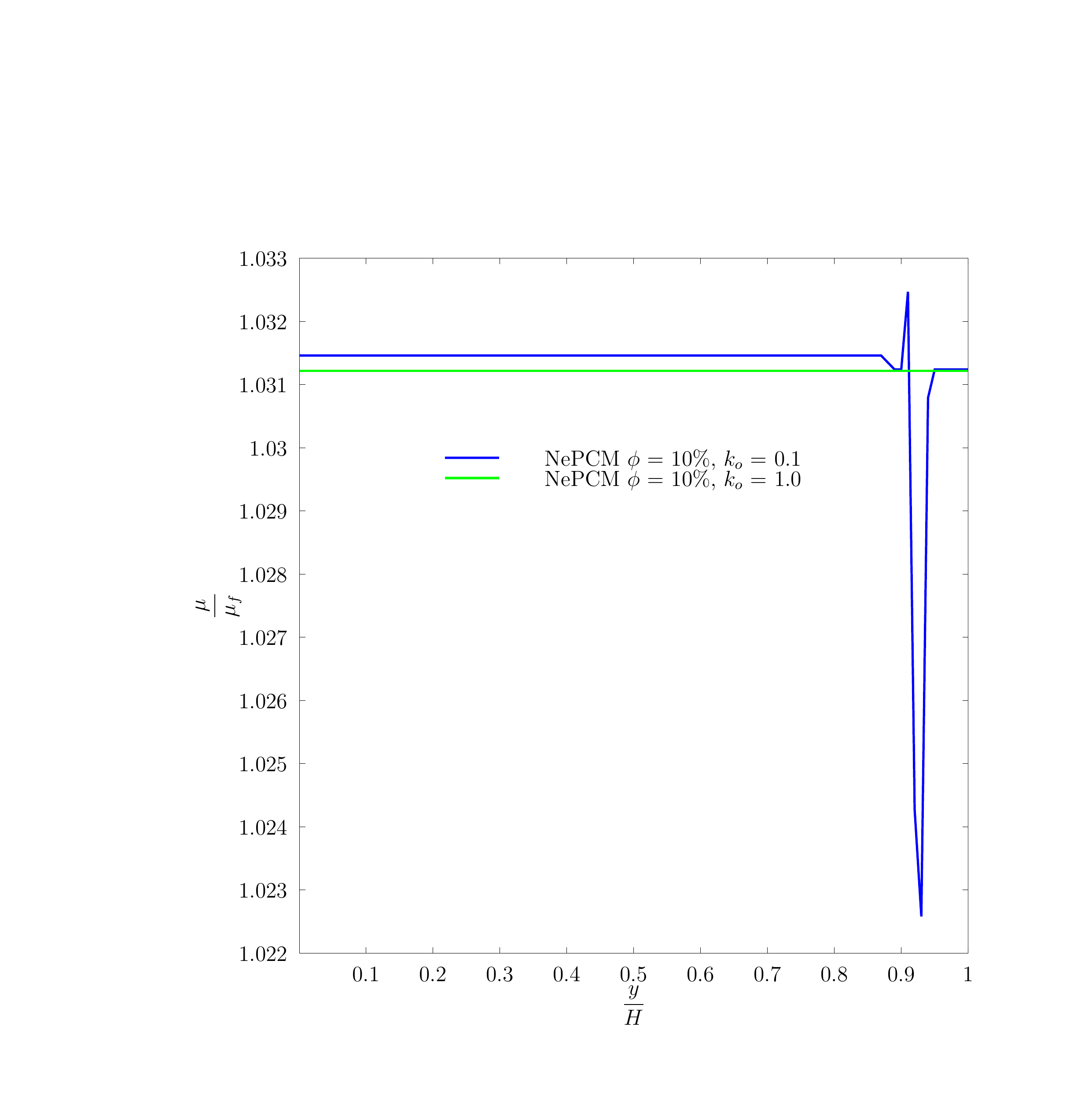}
\end{center}
\caption{The variation of the ratio between the viscosity  of the NePCM and that of pure water  along a vertical line that crosses the bottom, and top sides of the cavity at $\dfrac{x}{H}$ = 0.5, for the cases  of melting from the bottom for  $T_h$ = 300 K, $T_c$ = 270 K, for  NePCM with $\phi_w$ = 10\%, and different segregation coefficients\@. }
\label{Figviscousity_line_bootom_melting_different_k}
\end{figure}

We  also plot the viscosity ratio of NePcM to that of pure water for a vertical line extending from the bottom to the top walls located at $x/H$ = 0.5 for $k_o$ = 1.0 and 0.1, as depicted in Figure \ref{Figviscousity_line_bootom_melting_different_k}. A slight increase in the viscosity of the liquid NePCM is observed with $k_o$ = 0.1 compared to NePCM melted with $k_o$ = 1.0\@. This increase in viscosity is due to the rejection of particles during the melting process for the case of NePCM with $k_o$ = 0.1. The viscosity increase amounts to approximately \eval*{$\frac{1.03124-1.03122}{1.03124}\times 10^{2}$}\%, which is negligible and  highlights that the reduction in melting speed observed for NePCM with $k_o$ = 0.1 is attributable not to viscosity increase, but rather to thermosolutal convection.
Our simulations have conclusively demonstrated the significant role of thermosolutal convection in shaping the NePCM melting process and the corresponding flow fields. Furthermore, it is essential to note that thermosolutal convection represents a novel physical mechanism that must be considered to explain the degradation of NePCM observed in experiments. 

\subsubsection{Effect of mass fraction of particles}
\begin{figure}[p!]
\centering
\begin{subfigure}{.5\textwidth}
 \centering
 \caption{$\phi_w$ = 0.5\%}
  \includegraphics[width=.8\linewidth, trim =2.0cm 2.0cm 2.0cm 1.0cm,clip ]{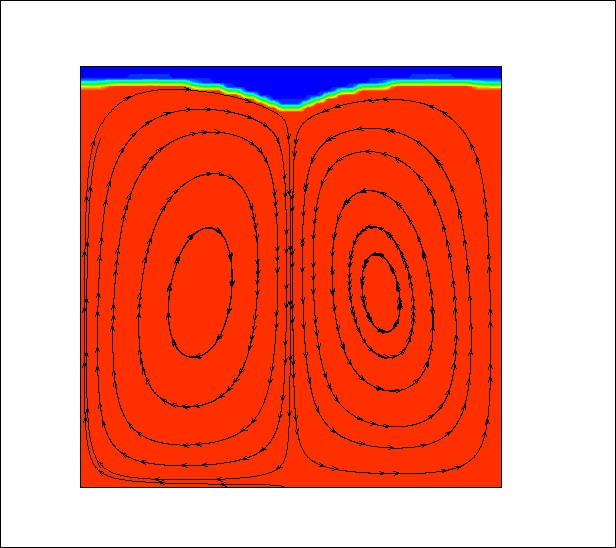}
\end{subfigure}%
\hfill
\begin{subfigure}{.5\textwidth}
  \centering
   \caption{$\phi_w$ = 5\%}
  \includegraphics[width=.8\linewidth,trim = 2.0cm 2.0cm 2.0cm 2.0cm,clip]{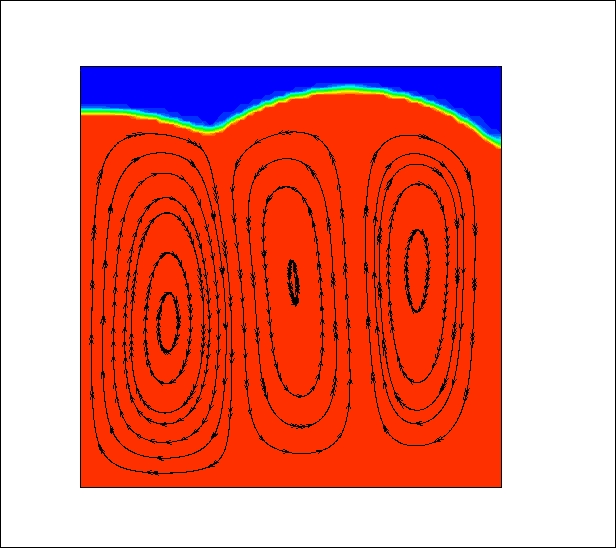}
\end{subfigure}%
\hfill
\begin{subfigure}{.5\textwidth}
  \centering
  \caption{$\phi_w$ = 10\%}
  \includegraphics[width=.8\linewidth,trim = 2.0cm 2.0cm 2.0cm 2.0cm,clip]{4.jpg}
\end{subfigure}
\bigskip

\begin{subfigure}{1.0\textwidth}
  \centering
  \includegraphics[width= .7\linewidth,trim = 0cm 0cm 0cm 0cm,clip]{liquid_fraction_top_bar.jpg}
\end{subfigure}
\caption{Development of the flow field (shown by the stream lines) superimposed on the contour of the liquid fraction   for  $T_h$ =300 K, $T_c$ = 270 K, for case  NePCM with different volume fractions, $t $ = 142 sec, and $k_o$ = 0.1\@.}
\label{Figliquidbottomdifferent_phi}
\end{figure}

We will investigate the effect of the mass fraction of the particles. We tested various mass loadings of particles while keeping the segregation coefficient at 0.1. Consequently, in all the considered NePCM samples, the ratio of the particles' mass fraction at the interface's solid and liquid sides remains equal. The liquid fraction distribution and the flow field during the late melting stage of the NePCM are depicted in Figure \ref{Figliquidbottomdifferent_phi}.

When \(\phi_w = 0.5\%\), the flow structure and the solid-liquid interface shape resemble waters. The melting process is predominantly controlled by thermal convection alone, given the reduced intensity of thermosolutal convection with lower particle mass fractions. With an increase in mass fraction to \(\phi_w = 5\%\), the flow transitions to a non-symmetric structure featuring three cells, accompanied by a non-symmetric solid-liquid interface shape. The flow structure and interface for \(\phi_w = 5\%\) closely resemble those of \(\phi_w = 10\%,\), underscoring the significance of particle mass fraction in shaping the flow structure and the liquid-solid interface shape during the NePCM melting process. This phenomenon, a novel contribution, is reported for the first time in the literature.

\section{Conclusions}
Our investigation employed numerical simulations to comprehensively analyze the melting dynamics of NePCM (Nano-enhanced Phase Change Materials), incorporating intricate considerations of particle mass transport\@. The current investigation  encompassed two distinct melting configurations: top-side and bottom-side melting scenarios. Through this comprehensive study, our findings yield the following key conclusions:
\begin{itemize}
\item Melting the NePCM from the top cavity side ensures the suppression of convection effects, presenting the sole configuration where thermal conductivity enhancement can be optimally utilized.
\item In scenarios where melting occurs from the bottom cavity side with pure water, the process is primarily controlled by thermal convection. However, in the case of NePCM melting, it is dominantly governed by thermosolutal convection.
\item The presence of thermosolutal convection results in a slower solid-liquid interface compared to pure water, consistent with numerous experimental observations. Notably, for a mass fraction of particles ($\phi_w$) of 10\%, the NePCM demonstrates approximately a 6\% deceleration in comparison to pure water.
\item For our investigations   the enhancement of thermal conductivity surpasses that of viscosity in all cases considered. Consequently, we exclude enhanced viscosity as the primary physical mechanism influencing the deceleration of the NePCM melting process.

\end{itemize}

\bibliography{22}
\bibliographystyle{nature}
\pagebreak

\begin{figure}[H]
\centering
\begin{tikzpicture}[scale=0.7, every node/.style={scale=0.7}]

\pgfmathsetseed{42}

\begin{scope}[xshift=0cm]
\draw[thick, double] (0,0) -- (0,8) -- (8,8) -- (8,0) -- cycle;

\node[below,font=\Large] at (4,0) {$T_c$ };
\node[above,font=\Large] at (4,8) {$T_h$ };

\fill[blue] (0,0) rectangle (8,4);
\fill[red] (0,4) rectangle (8,8);
 \node[green, overlay,  font=\fontsize{20}{24}\selectfont] at (4,6) { Liquid};
      \node[green, overlay,  font=\fontsize{20}{24}\selectfont] at (4,2) { Solid};
\foreach \i in {1,...,100} {
\pgfmathsetmacro\x{abs(rand*8)}
\pgfmathsetmacro\y{abs(rand*8)}
\filldraw[black] (\x,\y) circle (0.1);
}

\node[above, font=\Large] at (4,10) (cavitya) {(a)};
\end{scope}

\begin{scope}[xshift=10cm]
\draw[thick, double] (0,0) -- (0,8) -- (8,8) -- (8,0) -- cycle;

\node[below, font=\Large] at (4,0) {$T_c$};
\node[above,font=\Large] at (4,8) {$T_h$};
\

\fill[red] (0,0) rectangle (8,4);
\fill[blue] (0,4) rectangle (8,8);
 \node[green, overlay,  font=\fontsize{20}{24}\selectfont] at (4,6) { Solid};
      \node[green, overlay,  font=\fontsize{20}{24}\selectfont] at (4,2) { Liquid};

\foreach \i in {1,...,100} {
\pgfmathsetmacro\x{abs(rand*8)}
\pgfmathsetmacro\y{abs(rand*8)}
\filldraw[black] (\x,\y) circle (0.1);
}

\node[above, font=\Large] at (4,10) (cavitya) {(b)};
\end{scope}

\end{tikzpicture}
\caption{ The geometry under-consideration (a) melting from the top side (b) melting from the bottom side  }
\label{geomtery}
\end{figure}
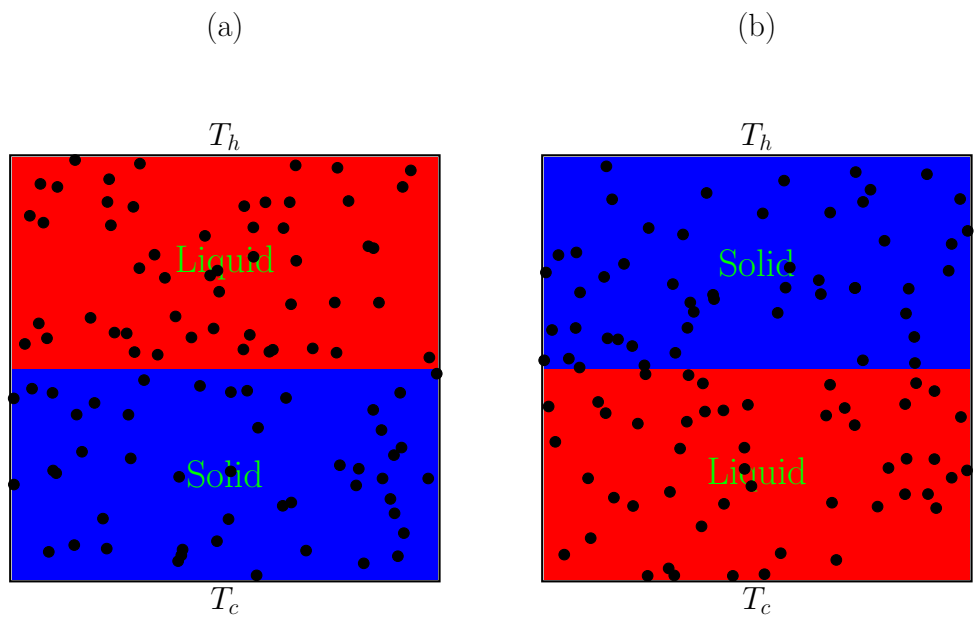
\begin{figure}[p!]
\centering
\begin{subfigure}{.5\textwidth}
 \centering
 \caption{t= 8 sec}
  \includegraphics[width=.8\linewidth, trim =2.0cm 2.0cm 2.0cm 2.0cm,clip ]{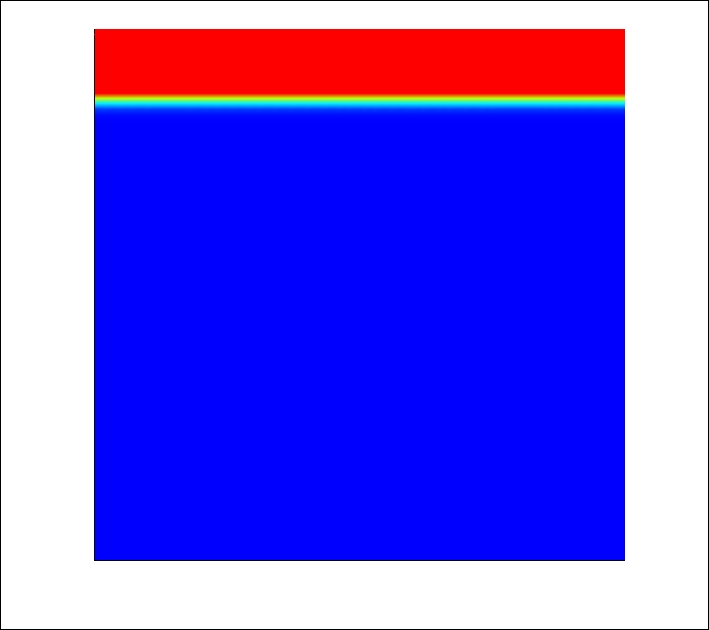}
\end{subfigure}%
\begin{subfigure}{.5\textwidth}
  \centering
  \caption{t = 50 sec}
  \includegraphics[width=.8\linewidth,trim = 2.0cm 2.0cm 2.0cm 2.0cm,clip]{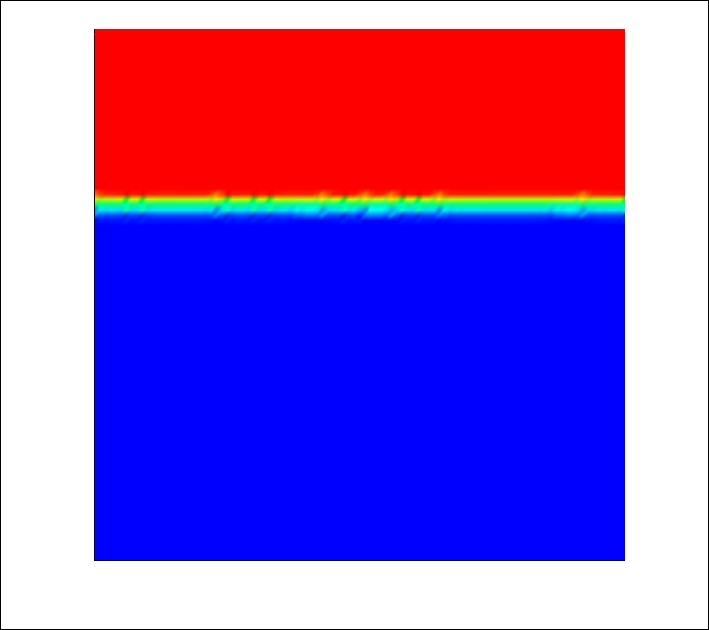} 
\end{subfigure}
\bigskip
\begin{subfigure}{.5\textwidth}
  \centering
   \caption{t = 140 sec}
  \includegraphics[width=.8\linewidth,trim = 2.0cm 2.0cm 2.0cm 2.0cm,clip]{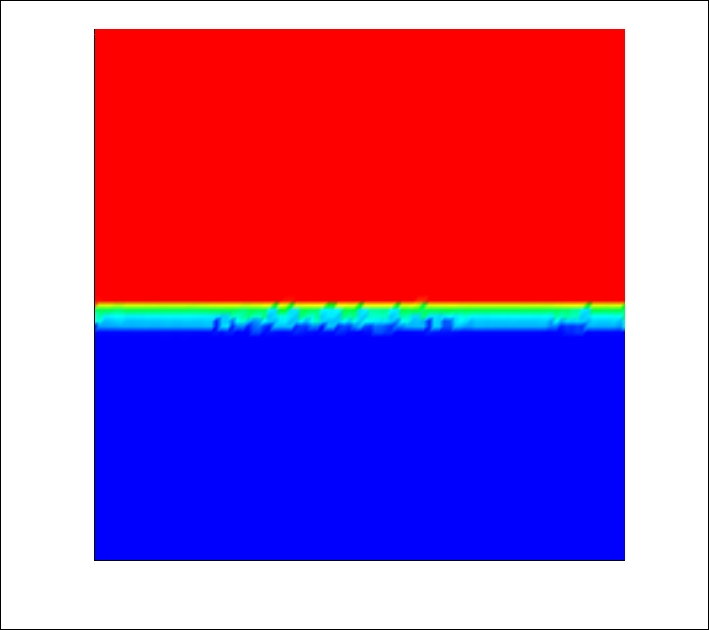}
\end{subfigure}%
\begin{subfigure}{.5\textwidth}
  \centering
  \caption{t = 200 sec}
  \includegraphics[width=.8\linewidth,trim = 2.0cm 2.0cm 2.0cm 2.0cm,clip]{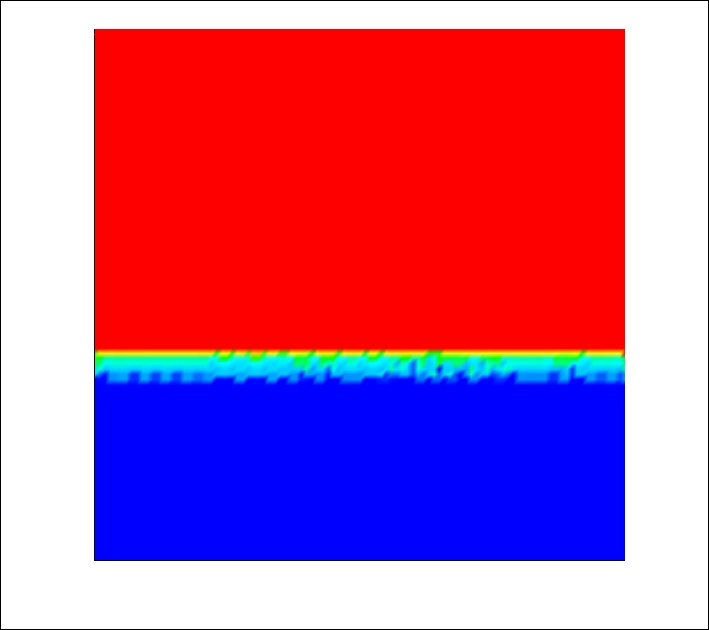}
\end{subfigure}
\bigskip
\begin{subfigure}{.5\textwidth}
  \centering
\includegraphics[width= 1.2\linewidth,trim = 0cm 0cm 0cm 0cm,clip]{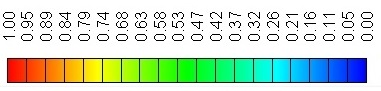}
\end{subfigure}
\caption{The evolution of liquid fraction profiles for the case of melting from the top for  $T_h$ = 290 K, $T_c$ = 270 K, and $\phi_w$ = 10\%.}
\label{Figliquidtopm}
\end{figure}



\begin{figure}[p!]
\centering
\begin{subfigure}{.5\textwidth}
 \centering
 \caption{t= 8 sec}
  \includegraphics[width=.8\linewidth, trim =2.0cm 2.0cm 2.0cm 1.0cm,clip ]{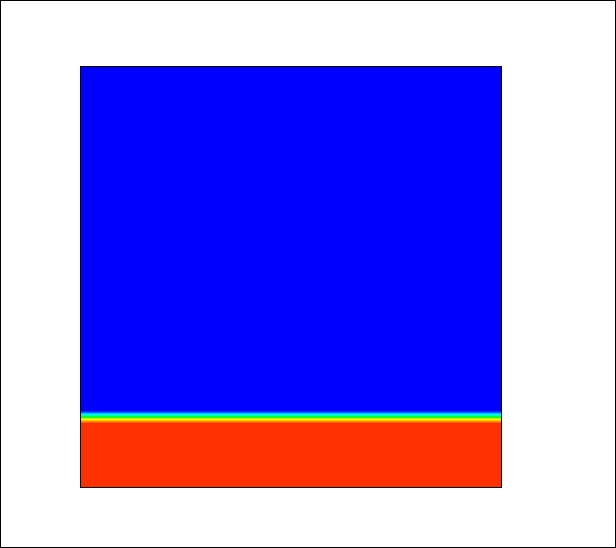}
\end{subfigure}%
\begin{subfigure}{.5\textwidth}
  \centering
  \caption{t = 52 sec}
  \includegraphics[width=.8\linewidth,trim = 2.0cm 2.0cm 2.0cm 2.0cm,clip]{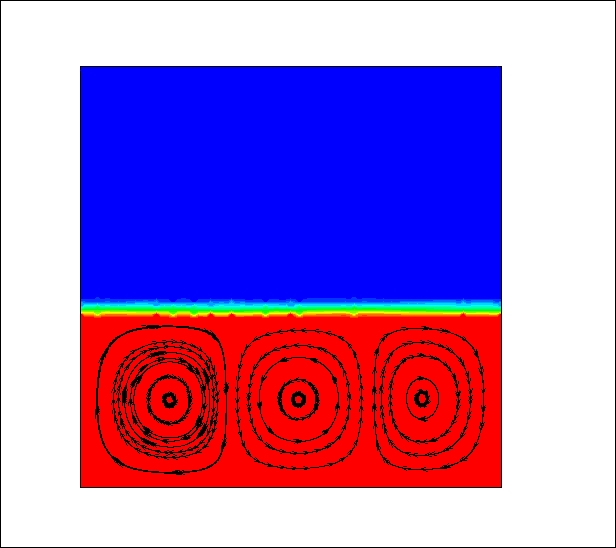}
\end{subfigure}
\bigskip
\begin{subfigure}{.5\textwidth}
  \centering
   \caption{t = 127 sec}
  \includegraphics[width=.8\linewidth,trim = 2.0cm 2.0cm 2.0cm 2.0cm,clip]{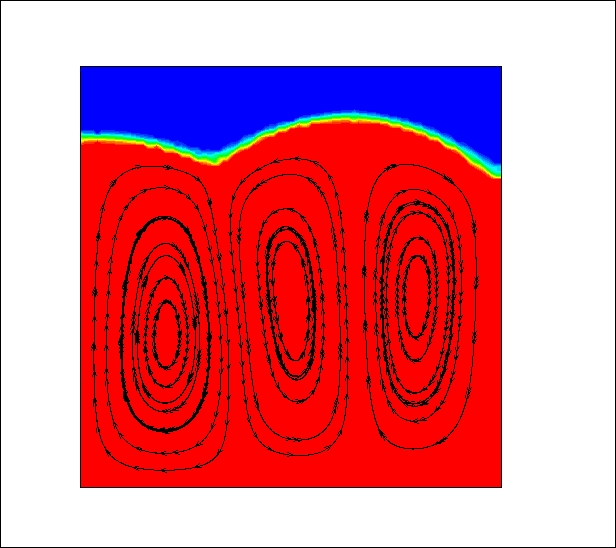}
\end{subfigure}%
\begin{subfigure}{.5\textwidth}
  \centering
  \caption{t = 143 sec}
  \includegraphics[width=.8\linewidth,trim = 2.0cm 2.0cm 2.0cm 2.0cm,clip]{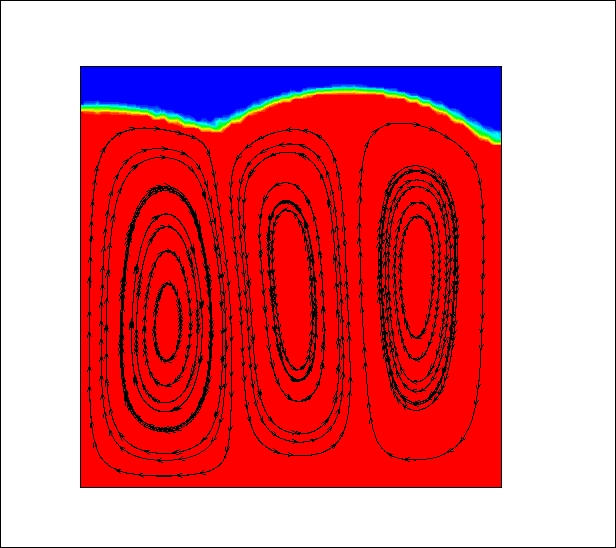}
\end{subfigure}
\bigskip
\begin{subfigure}{.5\textwidth}
  \centering
  \includegraphics[width= 1.2\linewidth,trim = 0cm 0cm 0cm 0cm,clip]{liquid_fraction_top_bar.jpg}
\end{subfigure}
\caption{Development of the flow field (shown by the stream lines) superimposed on the contour of the liquid fraction   for  $T_h$ =300 K, $T_c$ = 270 K, for case of NePCM with $\phi_w$ = 10\%\@.}
\label{Figliquidbottom3}
\end{figure}


\end{document}